\documentclass[8pt,a4paper,usenatbib]{mnras}
\usepackage[utf8]{inputenc}
\usepackage{amsmath}
\usepackage{amsfonts}
\usepackage{amssymb}
\usepackage{graphicx}
\usepackage{lmodern}
\usepackage{multicol}
\usepackage{hyperref}
\usepackage{verbatim}
\usepackage{float}
\usepackage{xspace}

\usepackage{ulem}
\usepackage[usenames]{color}

\usepackage{mathrsfs,bm}
\newcommand{\bcdot}{\ensuremath{%
  \mathchoice%
   {\mskip\thinmuskip\lower0.2ex\hbox{\scalebox{1.5}{$\cdot$}}\mskip\thinmuskip}}%
   {\mskip\thinmuskip\lower0.2ex\hbox{\scalebox{1.5}{$\cdot$}}\mskip\thinmuskip}%        
   {\lower0.3ex\hbox{\scalebox{1.2}{$\cdot$}}}%  
   {\lower0.3ex\hbox{\scalebox{1.2}{$\cdot$}}}%
}

\newcommand{\CR}{\rmn{cr}}
\renewcommand{\th}{\rmn{th}}
\newcommand{\de}{\mathrm{d}}
\newcommand{\di}{\partial}
\newcommand{\mach}{\mathcal{M}}

\renewcommand{\epsilon}{\varepsilon}

\newcommand{\AREPO}{{\sc arepo}\xspace}
\newcommand{\vecbf}{\mathbfit}
\newcommand{\bnabla}{\ensuremath{\boldsymbol{\nabla}}}

\newcommand{\dd}[2]{\frac{\mathrm{d} #1}{\mathrm{d} #2}}

\def\del#1{{}}

\author[Pais, Pfrommer, Ehlert \& Pakmor]{M. Pais$^{1}$, C. Pfrommer$^{1}$, K. Ehlert$^{1}$, and R. Pakmor$^{2}$
\\
$^{1}$Leibniz-Institut f\"{u}r Astrophysik Potsdam,  An der Sternwarte 16, 14482 Potsdam, Germany \\
$^{2}$Heidelberger Institut für Theoretische Studien, Schlo\ss -Wolfsbrunnenweg 35, 69118 Heidelberg, Germany  \\
}

\title{The effect of cosmic-ray acceleration on supernova blast wave dynamics}

\date{}

\begin{document}
\maketitle
\begin{abstract}
  Non-relativistic shocks accelerate ions to highly relativistic energies
  provided that the orientation of the magnetic field is closely aligned with
  the shock normal (quasi-parallel shock configuration). In contrast,
  quasi-perpendicular shocks do not efficiently accelerate ions. We model this
  obliquity-dependent acceleration process in a spherically expanding blast wave
  setup with the moving-mesh code {\sc arepo} for different magnetic field
  morphologies, ranging from homogeneous to turbulent configurations. A
  Sedov-Taylor explosion in a homogeneous magnetic field generates an oblate
  ellipsoidal shock surface due to the slower propagating blast wave in the
  direction of the magnetic field. This is because of the efficient cosmic ray
  (CR) production in the quasi-parallel polar cap regions, which softens the
  equation of state and increases the compressibility of the post-shock gas. We
  find that the solution remains self-similar because the ellipticity of the
  propagating blast wave stays constant in time. This enables us to derive an
  effective ratio of specific heats for a composite of thermal gas and CRs as a
  function of the maximum acceleration efficiency. We finally discuss the
  behavior of supernova remnants expanding into a turbulent magnetic field with
  varying coherence lengths. For a maximum CR acceleration efficiency of about
  15 per cent at quasi-parallel shocks (as suggested by kinetic plasma
  simulations), we find an average efficiency of about 5 per cent, independent
  of the assumed magnetic coherence length.
\end{abstract}


\section{Introduction}
Diffusive shock acceleration (DSA) is a universal process that operates at
strong, non-relativistic collisionless shocks and enables a small fraction of
particles impinging on the shock to gain more energy than the average particle
through multiple shock crossings \citep{1977ICRC...11..132A,1977DoSSR.234.1306K,
  1978MNRAS.182..147B,1978ApJ...221L..29B}.  The blast waves of supernova
remnants (SNRs) are the most likely sources of Galactic CRs
(\citealt{Neronov2017}; for extensive reviews, see \citealt{2008ARA&A..46...89R,
  2016RPPh...79d6901M}). There are other potential sources that might
contribute, including shocks associated with young star forming regions
\citep{2017arXiv171002803Y}, high-energy processes at the Galactic center
\citep{2016Natur.531..476H}, or shocks associated with a large-scale Galactic
wind that are driven by thermal or CR pressure gradients
\citep{2015MNRAS.453.3827S,2017ApJ...847L..13P}.

CR acceleration modifies the expansion history of a SNR shock due to the
additional CR pressure.  While the most energetic CRs escape the SNR upstream
and propagate into the interstellar medium (ISM), most of the CRs, by energy
content and by particle number, are swept downstream and end up in the interior
of the SNR \citep{2013MNRAS.431..415B} until they are eventually released to the
ISM when the SNR shell breaks into individual pieces as a result of
Rayleigh-Taylor instabilities that develop once the shock wave has sufficiently
slowed down. A self-similar Sedov-Taylor blast wave solution that accounts for
CR pressure was developed by \citet{1983ApJ...272..765C} and generalized to
include a CR spectrum and the maximum CR energy
\citep{2015MNRAS.447.2224B}. Those works demonstrate that the CR pressure
inevitably dominates the thermal pressure in the SNR interior even if only a
small fraction of the shock kinetic energy is converted to CRs. This is because
of the smaller ratio of specific heats of CRs ($\gamma_{\rmn{cr}}=4/3$) in
comparison to a thermal fluid ($\gamma_{\rmn{th}}=5/3$), which cause the thermal
pressure to decrease at a faster rate in comparison to CRs upon adiabatic
expansion.  Simulations of DSA at Sedov-Taylor blast waves confirmed that the
increased compressibility of the post-shock plasma due to the produced CRs
decreases the shock speed \citep[][for one- and three-dimensional simulations,
  respectively]{2011ApJ...734...85C,2017MNRAS.465.4500P}.

%\clearpage

However, these approaches missed one important plasma physics aspect of the
acceleration process: the orientation of the upstream magnetic field. Hybrid
particle-in-cell (PIC) simulations (with kinetic ions and fluid electrons) of
non-relativistic, large Mach number shocks demonstrated that DSA of ions
operates for quasi-parallel configurations (i.e., when the upstream magnetic
field is closely aligned with the shock normal), and becomes ineffective for
quasi-perpendicular shocks \citep{2014ApJ...783...91C}.  Ions that enter the
shock when the discontinuity is the steepest are specularly reflected by the
electrostatic shock potential and are injected into DSA
\citep{2015ApJ...798L..28C}.  Scattering of protons and electrons is mediated by
right-handed circularly polarized Alfv{\'e}n waves excited by the current of
energetic protons via non-resonant hybrid instability
\citep{2004MNRAS.353..550B}. After protons gained energy through a few
gyrocycles of shock drift acceleration (SDA), they participate in the DSA
process. On the contrary, after preheated via SDA, electrons are first
accelerated via a hybrid process that involves both SDA and Fermi-like
acceleration mediated by Bell waves, before they get injected into DSA
\citep{2015PhRvL.114h5003P}.

While quasi-perpendicular shocks are unable to accelerate protons, these
configurations can energize thermal electrons at the shock front via SDA. The
accelerated electrons are then reflected back upstream where their interaction
with the incoming flow generates oblique magnetic waves that are excited via the
firehose instability \citep{2014ApJ...794..153G, 2014ApJ...797...47G}. The
efficiency of electron injection is strongly modulated with the phase of the
shock reformation. Ion reflection off of the shock leads to electrostatic
Buneman modes in the shock foot, which provide first-stage electron energisation
through the shock-surfing acceleration mechanism \citep{2017ApJ...847...71B}.

In this work, we are studying magnetic obliquity-dependent acceleration of
protons at a strong, total energy conserving shock that is driven by a
point explosion \citep[similar in spirit to the analytic model
  by][]{2012MNRAS.419.1421B}. Hence, our setup models the Sedov-Taylor phase of
an expanding SNR and we examine how CR acceleration modifies its propagation
depending on the upstream properties of the magnetic field.  We emphasize that
we do not consider the pressure-driven snowplow phase of SNRs that begins $\sim2
\times 10^4$ years after the explosion and is characterized by radiative losses
of the shocked medium. The snowplough effect adiabatically compresses ambient
magnetic fields, which modifies the morphological appearance of the remnant
considerably \citep{2015A&A...584A..49V}.

This paper is organized as follows. In Section~\ref{sec:method} we present our
methodology, explain how we model magnetic obliquity-dependent CR shock
acceleration, and demonstrate the accuracy of our algorithm.  In
Section~\ref{sec:sedov} we present our Sedov-Taylor simulations with CR
acceleration: after deriving an analytical model on how the effective ratio of
specific heats depends on the CR acceleration efficiency, we show our blast wave
simulations with obliquity-dependent CR acceleration at homogeneous and
turbulent magnetic field geometries with varying correlation lengths.  In
Section~\ref{sec:conclusions} we summarize our main findings and conclude.  In
Appendix~\ref{sec:convergence} we assess numerical convergence of our algorithm.
In Appendix~\ref{sec:ST_analytics}, we numerically solve the system of equations
of a spherically symmetric gas flow to determine a relation between the
effective adiabatic index of the gas interior to the blast wave and the
self-similarity constant in the Sedov-Taylor solution. In
Appendix~\ref{sec:ellipsoid} we define the ellipsoidal reference frame that we
adopt for our oblate explosions and in Appendix~\ref{sec:dipole} we show our
results for obliquity dependent CR acceleration of a Sedov-Taylor explosion into
a dipole magnetic field configuration.

\begin{figure}
\label{fig:1}
\begin{center}
\includegraphics[width=0.48\textwidth]{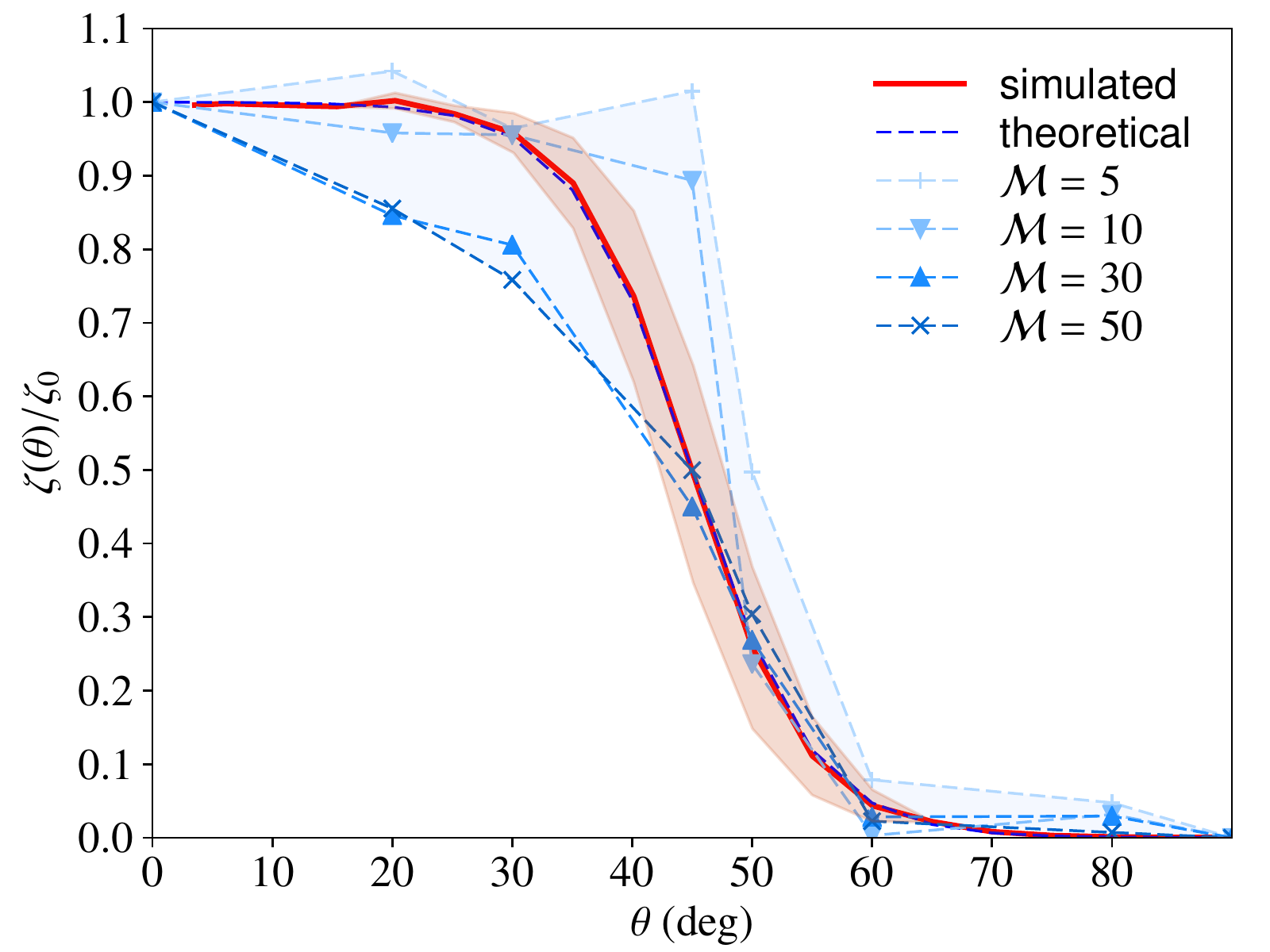} 
\end{center}
\caption{Functional dependence of the CR acceleration efficiency on the magnetic
  obliquity angle from hybrid PIC simulations of non-relativistic shocks
  \citep{2014ApJ...783...91C}. The blue points and dashed curves represent the
  results for different Mach numbers ranging from $\mach = 5$ to $50$,
  normalized to the maximum efficiency, respectively. The red curve represents
  our analytical model (equation~\ref{eq:efficiency}). All hybrid PIC
  simulations exhibit a sharp drop of the acceleration efficiency at the
  critical angle for shock acceleration, $\theta_{\mathrm{crit}} =
  45^\circ$. The coloured red region around our analytical model shows the
  accuracy of recovering this functional form in our 3D shock tube simulations
  (see Fig.~\ref{fig2:Shocktubes}).}
\label{fig1:efficiency}
\end{figure}

\section{Methodology}
\label{sec:method}

Here we present our methodology and explain the numerical algorithms to
implement magnetic obliquity dependent CR acceleration. We then validate our
implementation with shock tube simulations that exhibit homogeneous magnetic
fields. Finally, we lay down our procedure of setting up a turbulent magnetic
field that finds application in Sedov-Taylor explosions in
Sect.~\ref{sec:sedov}.

\begin{table*}
  \label{tab:1}
  \begin{tabular}{|c|c|c|c|c|c|c|c|c|c|c|}
    \hline 
    \hline
    $\theta$ & $\sqrt{\vecbf{B}^2}$ & $\rho_\mathrm{l}$ &  $\rho_\mathrm{r}$  & $P_\mathrm{l}$ &  $P_\mathrm{r}$ & $X^{\mathrm{cr}}_\mathrm{l}$  & $X^{\mathrm{cr}}_\mathrm{r}$ & $r_\rmn{c}$ & $\mach_1$ & $\gamma_{\mathrm{eff}}$ \\ 
    \hline 
    0$^{\circ}$& 10$^{-6}$ & 1 & 0.125 & 51.516 & 0.1 & 2 & 1 & 4.78 & 9.56 & 1.50\\ 
    45$^{\circ}$ & 10$^{-6}$ & 1 & 0.125 & 51.516 & 0.1 & 2 & 1 & 4.28 & 9.78 & 1.58\\ 
    90$^{\circ}$ & 10$^{-6}$ & 1 & 0.125 & 51.516 & 0.1 & 2 & 1 & 3.90 & 10.00 & 1.66\\ 
    \hline 
  \end{tabular}
  \centering
  \caption{Initial setup for the shock tubes shown in
    Fig. \ref{fig2:Shocktubes}. The columns show magnetic obliquity $\theta$,
    magnetic field strength $\sqrt{\vecbf{B}^2}$, initial mass density on the
    left- and right-hand sides, $\rho_\mathrm{l}$ and $\rho_\mathrm{r}$, total
    pressure on the left- and right-hand sides, $P_\mathrm{l}$ and
    $P_\mathrm{r}$, CR-to-thermal pressure ratio on the left- and right-hand
    sides, $X^{\mathrm{cr}}_\mathrm{l}$ and $X^{\mathrm{cr}}_\mathrm{r}$. The
    resulting shock has a compression ratio, $r_\rmn{c}$, Mach number $\mach_1$, and
    effective adiabatic index in the post-shock regime
    $\gamma_{\mathrm{eff}}$ (see equation~\ref{eq:gamma_i}). }
\end{table*}

\subsection{Simulation method}

All simulations in this paper are carried out with the massively parallel code
\AREPO \citep{2010MNRAS.401..791S} in which the gas physics is calculated on a
moving Voronoi mesh, using an improved second-order hydrodynamic scheme with
the least-squares-fit gradient estimates and a Runge-Kutta time integration
\citep{2016MNRAS.455.1134P}. We calculate the fluxes across the moving interface
from the reconstructed primitive variables using the HLLD Riemann solver
\citep{2005JCoPh.208..315M}.  We follow the equations of ideal MHD coupled to a
second, CR fluid using cell-centred magnetic fields and the
\citet{1999JCoPh.154..284P} scheme for divergence control
\citep{2011MNRAS.418.1392P, 2013MNRAS.432..176P}.

We model the relativistic CR fluid with an adiabatic index $\gamma = 4/3$ and
account for diffusive shock acceleration of CRs at resolved shocks in the
computational domain, following a novel scheme \citep{2017MNRAS.465.4500P}.
In our subgrid model for CR acceleration, we assume that diffusive shock
  acceleration operates efficiently provided there are favorable conditions
  (e.g., Sect.~\ref{sec:obliquity}). This can be realized in the physical
  scenario, in which the current associated with the forward streaming CRs
  excites the non-resonant hybrid instability \citep{2004MNRAS.353..550B}. This
  leads to exponential growth of magnetic fluctuations until the instability
  saturates at equipartition with the kinetic energy flux. This also implies
  efficient pitch angle scattering of CRs so that they approach the Bohm limit
  of diffusion. In such a situation, we can calculate the CR precursor length
  for the pressure-carrying protons between 1 to 10 GeV,
  \begin{align}
    \label{eq:direction}
    L_\rmn{prec} &\sim\sqrt{\kappa_\rmn{Bohm}t}\\
    &\sim0.001\,\rmn{pc}
    \left(\frac{pc}{10\,\rmn{GeV}}\right)^{1/2}
    \left(\frac{B}{100\,\umu\rmn{G}}\right)^{-1/2}
    \left(\frac{t}{10^3\,\rmn{yr}}\right)^{1/2}.\nonumber
  \end{align}
  The CR precursor length only raises to 0.1 pc for TeV CRs gyrating in $\umu$G
  fields, which is still smaller than the numerical resolution $\Delta
  x=0.125$~pc of our simulations and thus unresolved (assuming a typical box
  size of 25~pc and $200^3$ grid cells for SNR simulations). Hence, in the
interest of a transparent setup, we only model the dominant advective CR
transport and neglect CR diffusion and streaming.

Similarly, we only account for adiabatic CR losses and neglect non-adiabatic CR
losses such as Coulomb, hadronic and Alfv\'en-wave losses. In particular, we
  neglect the small effect of energy loss from the blast wave due to CRs
  escaping upstream. This effect softens the Sedov-Taylor solution from $r_s
  \propto t^{0.4}$ to $r_s \propto t^{0.39}$ \citep{2015MNRAS.447.2224B}. That
  calculation assumed a momentum spectrum of $p^{-4}$, which provides an upper
  limit to the energy contribution of escaping high-energy CRs. For
  observationally inspired softer spectra, the softening of the Sedov-Taylor
  solution becomes even smaller, thus justifying our neglect.

To localize shocks and their up- and downstream properties during the run time
of the simulation, we adopt the method by \citet{2015MNRAS.446.3992S} that is
solely based on local cell-based criteria. The method identifies the direction
of shock propagation in Voronoi cells that exhibit a converging flow with the
negative gradient of the pseudo-temperature that is defined as
\begin{equation}
  \label{eq:temp}
  k T^* = \frac{P}{n} = \frac{\mu m_\rmn{p} (P_\th + P_\CR)}{\rho},
\end{equation}
where $n$ is the number density, $m_\rmn{p}$ is the proton rest mass, and $\mu$ is the
mean molecular weight. Hence, the shock normal is given by
\begin{equation}
  \label{eq:direction}
  \bm{\hat{n}}_\rmn{s} = -\frac{\bnabla T^*}{\left|\bnabla T^*\right|}.
\end{equation}

Voronoi cells with shocks are identified with (i) a maximally converging flow
along the direction of shock propagation, while (ii) spurious shocks such as
tangential discontinuities and contacts are filtered out, and (iii) the method
provides a safeguard against labelling numerical noise as physical shocks.  In
particular, in our study the magnetic field is dynamically irrelevant at the
shock, such that the non-MHD jump conditions are valid.  Typically, shocks in
\AREPO are numerically broadened to a width of two to three cells. By extending
the stencil of the shock cell into the true pre- and post-shock regime, we
determine the Mach number and dissipated energy of the shock. This enables us to
inject a pre-determined energy fraction into our CR fluid into those Voronoi
cells that exhibit a shock and into the adjacent post-shock cells
\citep[see][for more details]{2017MNRAS.465.4500P}.

\subsection{Obliquity-dependent CR acceleration}
\label{sec:obliquity}

\begin{figure*}
\centering
\includegraphics[scale=0.4]{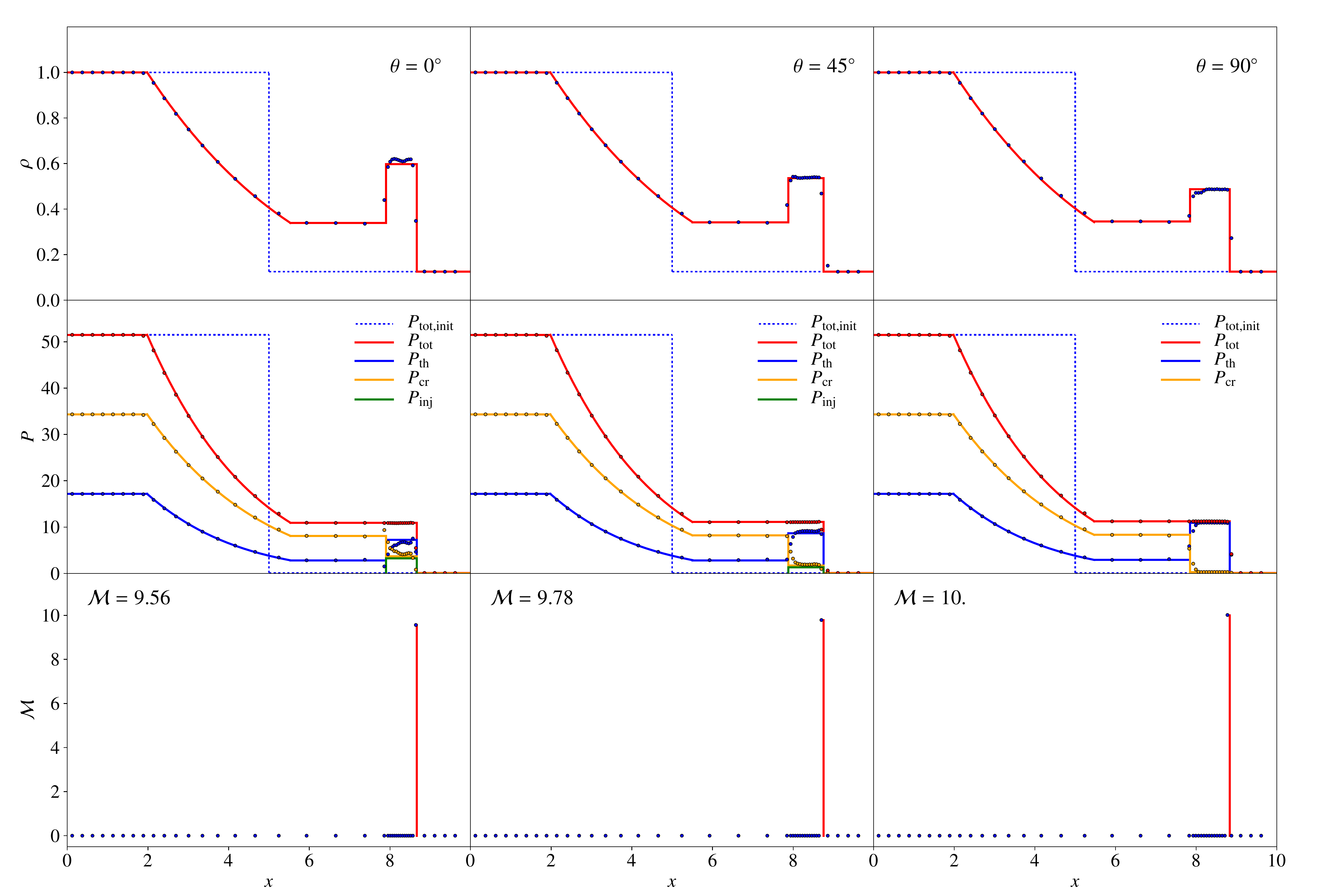} 
\caption{Shock-tube tests for different magnetic field orientations that account
  for pre-existing and freshly accelerated CRs (solid lines show the analytic
  solution of \citet{2017MNRAS.465.4500P}, data points show mean simulation
  values, each averaged over 250 Voronoi cells). Shown are 3D simulations at
  time $t=0.35$. For each simulation we show density $\rho$, pressure $P$ and
  Mach number (top to bottom). The left panels show the simulation with a
  parallel magnetic shock configuration ($\theta=0$), where the acceleration
  efficiency $\zeta$ is maximized. The middle panels adopt an oblique
  configuration with $\theta=45^\circ$. We notice that the post-shock gas is
  less dense in comparison to the parallel case due to the smaller CR pressure
  as a result of a lower acceleration efficiency. In the right column, we show a
  perpendicular magnetic configuration ($\theta=90^\circ$), for which CR
  acceleration is completely inefficient. In this case, the shock propagates
  fastest (i.e., with the largest Mach number $\mach$) of all three cases for
  otherwise identical initial conditions because of the absence of freshly
  injected CRs in the post-shock gas, which results in the hardest adiabatic
  index of $\gamma=5/3$ in the post-shock regime.}
\label{fig2:Shocktubes}
\end{figure*}

We adopt the following relation between the injected CR energy, $\Delta
E_{\mathrm{cr}}$ and the dissipated energy at the shock, $E_{\mathrm{diss}}$,
\begin{equation}
\label{deltaE}
\Delta E_{\mathrm{cr}} = \zeta(\mach_1, \theta) E_{\mathrm{diss}}.
\end{equation}
The injection efficiency $\zeta(\mach_1, \theta)$ depends on the shock Mach
number, $\mach_1=v_\rmn{s}/c_1$ (i.e., the shock speed in units of the pre-shock
sound speed, $c_1$) and the {\it upstream} magnetic obliquity, $\theta$, defined
as the angle between the normal to the shock front, $\vecbf{n}_\rmn{s}$, and the
direction of the magnetic field, $\bm{\hat{b}}=\vecbf{B}/|\vecbf{B}|$:
\begin{equation}
\cos\theta =  \bm{\hat{n}}_\rmn{s} \bcdot \bm{\hat{b}}.
\end{equation}
Since the physics does not depend on the actual direction of the unit vectors
$\bm{\hat{n}}_\rmn{s}$ and $\bm{\hat{b}}$ (i.e., whether the vectors point in the same
quadrant or not), we re-define the magnetic obliquity via
\begin{equation}
  \theta = \arccos(|\cos\theta|)
\end{equation}
In practice, for every shocked cell we collect the magnetic obliquity in the
corresponding pre-shock region and communicate it to the shocked cell.

We calibrate $\zeta(\mach_1, \theta)$ with hybrid PIC simulations performed by
\citet{2014ApJ...783...91C}. The authors find that DSA of ions is very efficient
at quasi-parallel shocks, producing non-thermal ion spectra with the expected
universal power-law distribution in momenta equal to $p^{-4}$. At very oblique
shocks, ions can be accelerated via shock drift acceleration, but they only gain
a factor of a few in momentum, and their maximum energy does not increase with
time. In this paper, we only consider strong shocks (i.e., $\mach_1 \gg 1$) for
which the injection efficiency saturates to a maximum value, $\zeta_0$. The
saturation happens for shocks with $\mach \gtrsim 30$, according to
\citet{2013ApJ...764...95K}.

Hence, we only need to model the scaling of the injection efficiency with
magnetic obliquity, which is shown in Fig.~\ref{fig:1} for different shock
strengths. All simulated curves of the injection efficiency (light blue curves
in Fig.~\ref{fig:1}) show a similar qualitative behavior: saturation at
quasi-parallel shocks, a steep decline at the threshold obliquity of
$\theta_{\mathrm{crit}}\simeq45^\circ$, and leveling off at zero for
quasi-perpendicular shocks.  However, at a given magnetic obliquity, the
function $\zeta(\theta,\mach)$ is not always monotonically rising with Mach
number and shows substantial scatter \citep[see Fig.~3
  of][]{2014ApJ...783...91C}. Hence, we decided to capture the qualitative
behavior of all four curves of the normalized injection efficiency,
$\zeta(\theta,\mach)/\zeta_0$, for different shock strengths with the following
functional form:
\begin{equation}
  \label{eq:efficiency}
  \zeta(\theta) \simeq \dfrac{\zeta_0}{2}
  \left[ \tanh\left( \dfrac{\theta_{\mathrm{crit}}-\theta}{\delta} \right) + 1 \right].
\end{equation}
We adopted a threshold obliquity of $\theta_{\mathrm{crit}}=\pi/4$ and a shape
parameter of $\delta = \pi/18$ (red curve in Fig.~\ref{fig:1}).  Hybrid PIC
simulations by \citet{2014ApJ...783...91C} demonstrate that the CR ion
acceleration efficiency saturates for large Mach numbers at a value of
$\zeta_0\simeq0.15$. In our simulations however, we adopt a maximum acceleration
efficiency of $\zeta_0=0.5$ to amplify the (dynamical) effects of CR
acceleration. We checked that reducing the acceleration efficiency to realistic
values results in qualitatively similar effects, albeit with a smaller amplitude. 

\begin{figure*}
\centering
\includegraphics[scale=0.55]{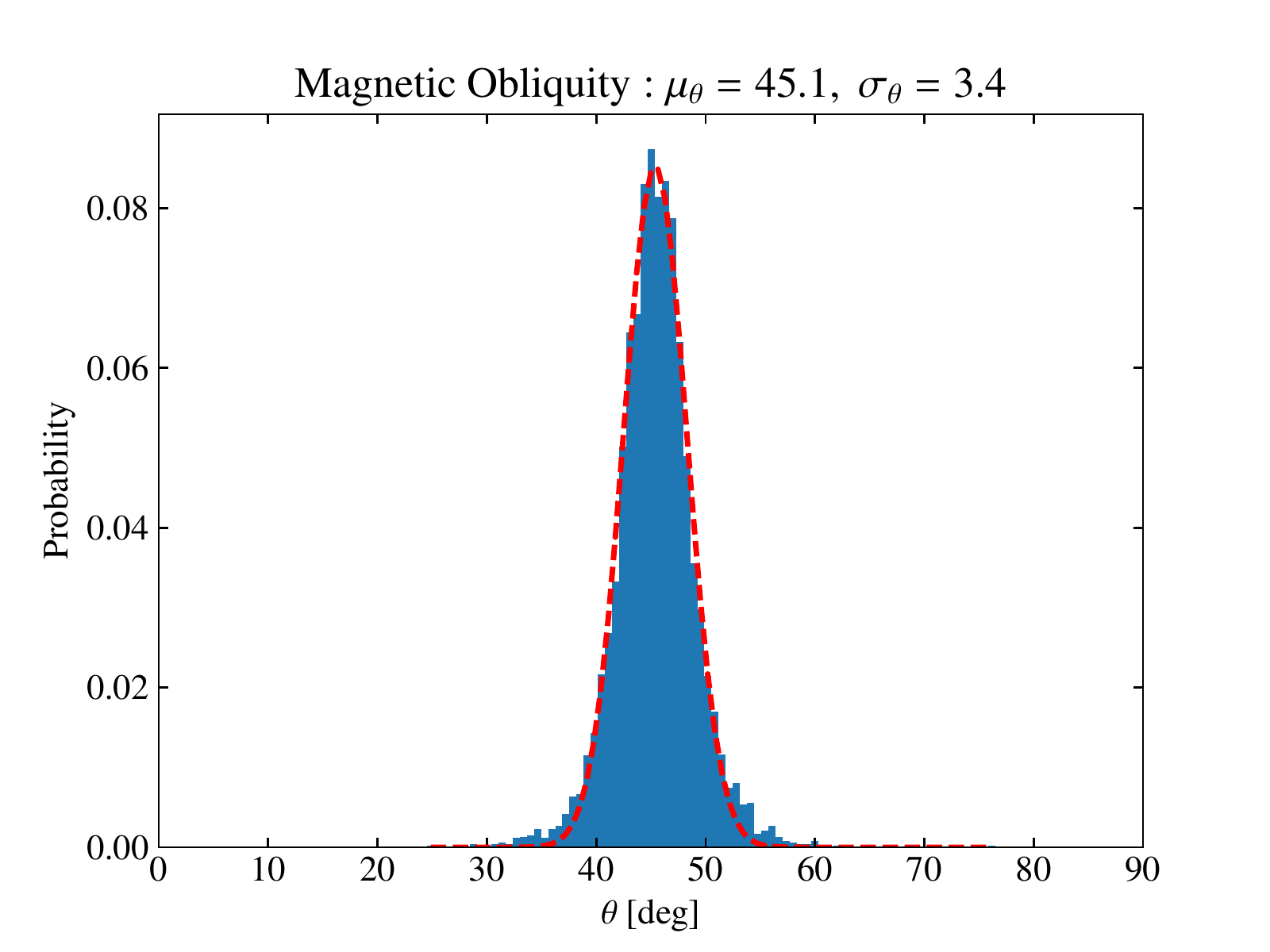} 
\includegraphics[scale=0.5]{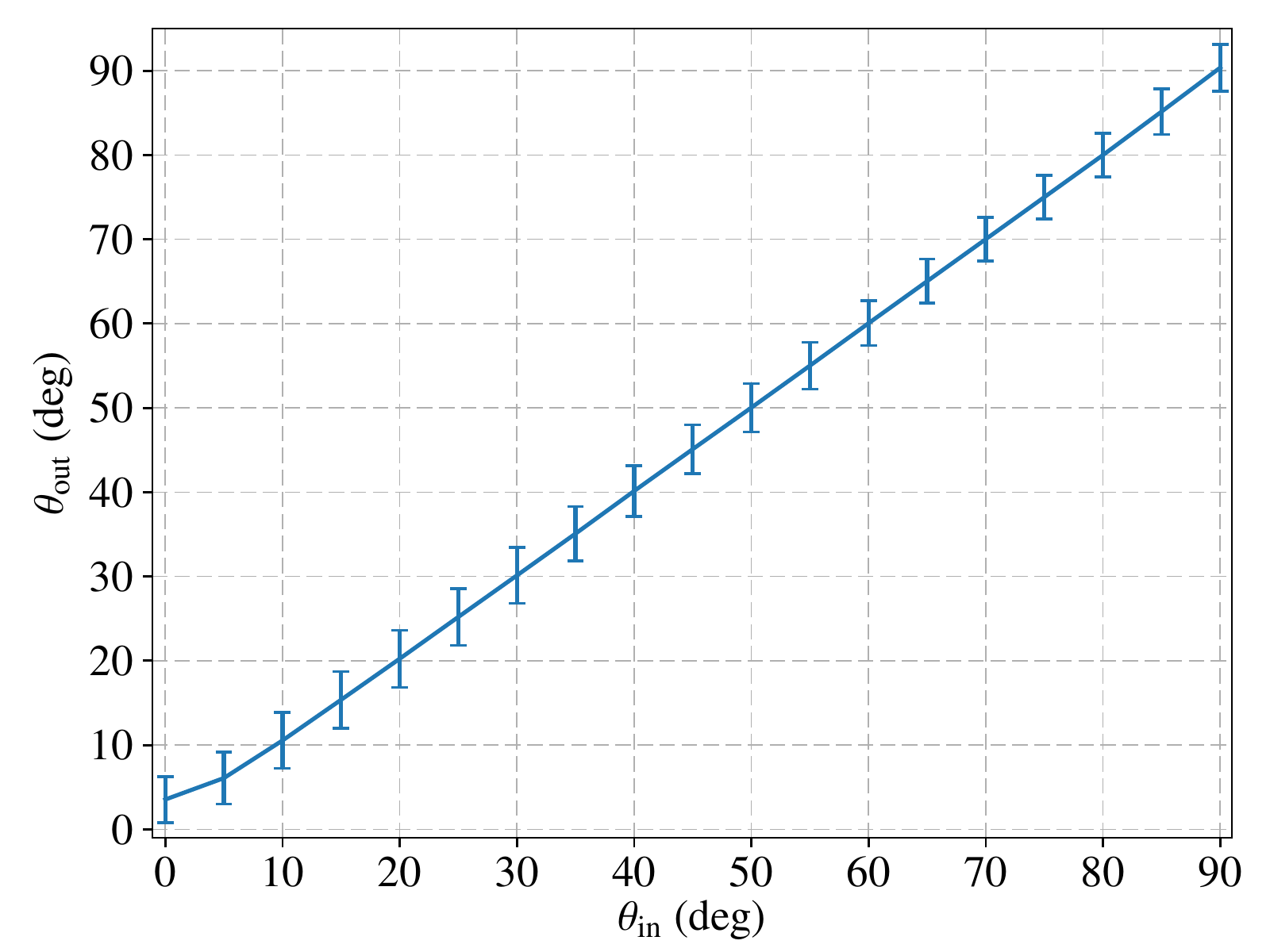} 
\caption{Left: probability distribution function (PDF) of the magnetic obliquity
  for a shock-tube simulation with initial obliquity of $45^\circ$. To increase
  the statistics, we use obliquity values of all shocked cells for 40
  equally-spaced snapshots of that simulation and fit the mean and variance of a
  Gaussian distribution (red dashed) to the PDF (blue histogram).  Right:
  comparison between the intrinsic ($\theta_\rmn{in}$) and recovered magnetic
  obliquity ($\theta_\rmn{out}$) for different shock-tube simulations. We
  recover the initial obliquity with an accuracy of $\sigma_\theta=3.4^\circ$
  for all simulations except for the case $\theta=0$ (which is however of little
  practical relevance due to the very similar shock acceleration efficiencies in
  this regime). }
\label{fig3:statistics}
\end{figure*}

\subsection{Code validation with shock tubes}

To validate our implementation and to test the correctness of our obliquity
dependent shock acceleration algorithm, we performed several Riemann shock-tube
simulations with different orientations of the magnetic field. A solution to the
shock-tube problem with accelerated CRs is derived analytically in
\citet{2017MNRAS.465.4500P} for a purely thermal gas and for a composite of
thermal gas and pre-existing CRs. In the limit of weak background magnetic
fields the solutions proposed in \citet{2017MNRAS.465.4500P} are still
applicable. We simulate three-dimensional (3D) shock tubes with initially $10^4$
cells in a box of dimension $10\times1\times1$. The initial Voronoi mesh is
generated by randomly distributing mesh-generating points in the simulation box  
and relaxing the mesh via Lloyd's algorithm \citeyearpar{journals/tit/Lloyd82} 
to obtain a glass-like configuration. All other initial parameters are laid down 
in Table~\ref{tab:1}.

In Fig.~\ref{fig2:Shocktubes}, we show three simulations with characteristically
different magnetic obliquities, $\theta = 0^\circ, 45^\circ$, and $ 90^\circ$.
Our choice of a larger total pressure on the left-hand side (with the tube
initially at rest), implies a rightwards moving shock, which is followed by a
contact discontinuity, as well as a leftwards moving rarefaction wave.  We show
mean simulation values of density, pressure and Mach number, each averaged over
250 Voronoi cells to ensure an identical Poisson error per bin and to
demonstrate the change of volume at the shock and over the rarefaction wave as a
result of the moving-mesh nature of \AREPO.

Changing the orientation of the magnetic field from quasi-parallel
($\theta\gtrsim0^\circ$) to quasi-perpendicular geometries
($\theta\lesssim90^\circ$), the acceleration process becomes less and less
efficient (as manifested from the fractions of post-shock CR pressure, see
second row in Fig.~\ref{fig2:Shocktubes}). In the case of $\theta = 90^\circ$,
CR acceleration is absent and the purely thermal case is restored with a
compression ratio of $r = 3.9$ (for the adopted initial conditions).  In the
first column of Fig.~\ref{fig2:Shocktubes} ($\theta = 0^\circ$), we see an
increased compressibility of the post-shock gas over the thermal case due to the
abundantly produced CRs, which yields a shock compression ratio of
$r=4.78$. Because of mass conservation, the shock cannot advance as fast in
comparison to the purely thermal case and the Mach number is accordingly lower.

Our implementation records the magnetic obliquity in the upstream of the shocked
Voronoi cells.  In the left panel of Figure \ref{fig3:statistics}, we present
the probability distribution function for $\theta$ for the intermediate case
$\theta = 45^\circ$.  To improve our statistics, we used 40 different
snapshots. We find normally distributed obliquity values around the expected
value, with a standard deviation of $3.4^\circ$. We repeated the experiment for 18
simulations with an input obliquity that differed by $5^\circ$ from the
preceding simulation.  The correspondence between injected angles
$\theta_\mathrm{in}$ and simulated angles $\theta_{\mathrm{out}}$ of shocked
Voronoi cells becomes apparent in the right panel of Figure
\ref{fig3:statistics}, with a 1-sigma accuracy of $3.4^\circ$. This accuracy is
numerically converged as we show in Appendix~\ref{sec:convergence}.  We only
observe a small numerical deviation at small obliquities
$\theta_\mathrm{in}<3^\circ$.  However, the resulting acceleration efficiency is
not affected due to the constant efficiency at quasi-parallel shocks.

\subsection{Turbulent magnetic fields}
\label{sec:turb_ICs}

In order to generate turbulent magnetic fields with an average value $\langle
\vecbf{B} \rangle =\mathbf{0}$ but $\langle \sqrt{\vecbf{B}^2}\rangle \neq 0$,
we adopt a magnetic power spectrum of Kolmogorov type and scale the field
strength to an average plasma beta factor of unity.  The three components of the
magnetic field $B_i$ ($i\in\{1,2,3\}$) are treated independently to ensure that
the final distribution of $\vecbf{B}(\vecbf{x})$ has a random phase.  To
proceed, we assume Gaussian-distributed field components that follow a
one-dimensional power spectrum $P_i(k)$, defined as $P_i(k) \propto k^2
|\tilde{B}_i (k)|^2 $, of the form
\begin{equation}
  |\tilde{B}_i (k)|^2 =
  \begin{cases} A, & k < k_{\mathrm{inj}}, \\
    A \left( \dfrac{k}{k_{\mathrm{inj}}}\right)^{-11/3}, & k_{\mathrm{inj}} \leq k,
  \end{cases}
 \end{equation} 
where $A$ is normalization constant, $k = |\vecbf{k}|$, and $k_{\mathrm{inj}}$
is the injection scale of the field. Modes on larger scales
($k<k_{\mathrm{inj}}$) follow a white noise distribution and modes with
$k>k_{\mathrm{inj}}$ obey a Kolmogorov power spectrum.    For
each magnetic field component, we set up a complex field such that
\begin{equation}
\label{reB}
      [\Re(\tilde{B}_i (\bm{k})), \Im(\tilde{B}_i (\bm{k}))]
      = [G_1(X_1, X_2) \sigma_k, G_2(X_1, X_2)  \sigma_k],
\end{equation}
where $G_i$ ($i\in\{1,2\}$) is a distribution of uniform random deviates $X_1$
and $X_2$ that returns Gaussian-distributed values. We set the corresponding
standard deviation $\sigma_k$ to $\tilde{B}_i$ for every value of $k$. We
normalize the spectrum to the desired variance of the magnetic field components
in real space, $\sigma_B$ using Parseval’s theorem,
\begin{equation}
\sigma_B^2 = \dfrac{1}{N^2} \sum_k^N |\tilde{B}_i (k)|^2.
\end{equation}
We then subtract the radial field component in $k$ space to fulfill the
constraint $\mathrm{div}\vecbf{B}=0$, via
\begin{equation}
  \tilde{\vecbf{B}}_k \to
  \tilde{\vecbf{B}}_k - \bm{\hat{k}}  (\bm{\hat{k}} \bcdot \tilde{\vecbf{B}}_k).
\end{equation}
Applying an inverse fast Fourier transform to $\tilde{\vecbf{B}}_k$ and
re-scaling the magnetic field to the desired average magnetic-to-thermal
pressure ratio yields a turbulent magnetic field distribution. To ensure
pressure equilibrium in the initial conditions, we adopt temperature
fluctuations of the form $n k_\rmn{B}\delta T=-\delta \vecbf{B}^2/(8\pi)$. This
setup does not balance the magnetic tension force. The resulting turbulent
motions have a small amplitude in comparison to the velocity of the expanding
blast wave so that to good approximation, the ambient gas can be considered
frozen and does not contribute to the dynamics.

\section{Sedov-Taylor explosions}
\label{sec:sedov}

In order to understand the non-thermal properties of supernova remnants, we
model the explosion as an expanding Sedov-Taylor blast wave in the magnetised
interstellar medium.  After deriving an analytical solution for the Sedov-Taylor
problem in the presence of CR acceleration with an arbitrary shock acceleration
efficiency, we study magnetic obliquity dependent CR acceleration in homogeneous
and turbulent fields with varying coherence scales and formulate an analytic
theory that enables us to obtain the average CR efficiency.

\begin{figure}
\centering
\includegraphics[scale=0.57]{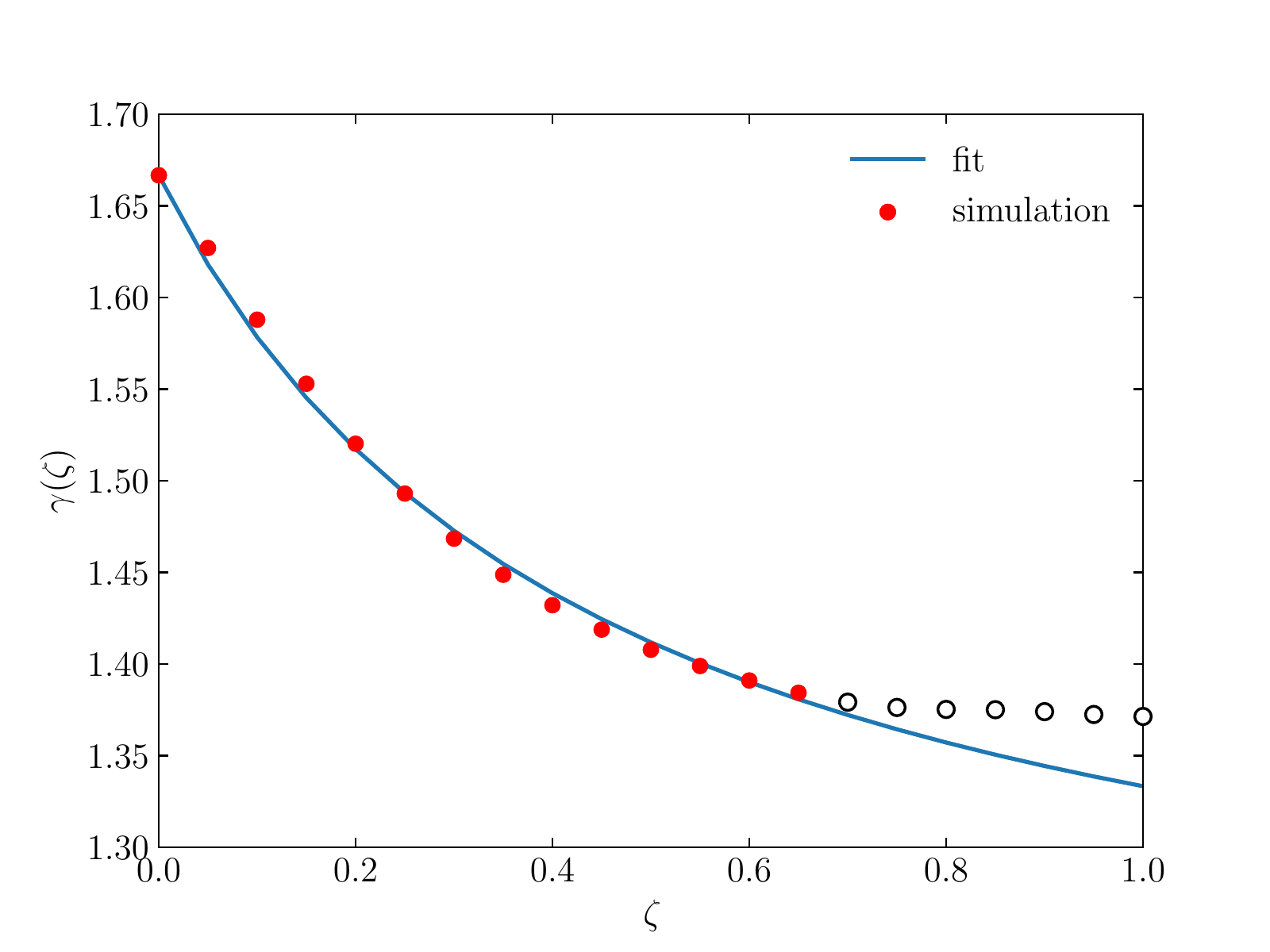} 
\caption{Effective ratio of specific heats $\gamma$ as a function of CR
  acceleration efficiency $\zeta$ for the Sedov-Taylor blast wave (ignoring the
  obliquity dependence of CR acceleration). The simulations (dots) and
  analytical fit (solid, equation~\ref{eq:alpha2}) interconnect the thermal gas
  case without CR acceleration (${\zeta=0}$) and the opposite extreme of a
  (hypothetical) 100\% efficient acceleration process, which yields a fully
  relativistic gas in the post-shock region ($\gamma=4/3$). The simulations do
  not reproduce the limit $\gamma\to4/3$ for $\zeta\to1$ due to residual
  thermalization.}
\label{fig4:mach}
\end{figure}

\subsection{Analytical solution with CR acceleration}
\label{sec:analytics}

\begin{table}
 \begin{center}
\begin{tabular}{c|c|c}
 \textbf{Parameter} & \textbf{Value} & \textbf{Approximation} \\ 
 \hline
 $a$ & 1.185 & 32/27\\
$b = 20/3 - 9a + 3 a^2 $ & 0.214 & 52/243 \\ 
$c = 4-3a$ & 0.445 & 4/9\\ 
\end{tabular}
\end{center}
 \caption{Best-fit parameters of the effective ratio of specific heats as a
   function of CR acceleration efficiency, $\gamma(\zeta)$, of
   equation~(\ref{eq:alpha2}) for the Sedov-Taylor blast-wave problem. The fit and the
   simulated points are shown in Fig.~\ref{fig4:mach}}
    \label{tab:2}
\end{table}

First, we derive analytical exact solutions of the Sedov-Taylor blast-wave
problem with CR acceleration without an obliquity dependent efficiency. If a
substantial fraction of the dissipated energy is converted into CRs, this alters
the effective adiabatic index $\gamma_{\mathrm{eff}}$ that is defined as the logarithmic
derivative of the total pressure with respect to density at constant entropy $s$:
\begin{equation}
  \label{eq:gamma_i}
  \gamma_{\rmn{eff}} \equiv \left.\dd{\ln (P_\CR+P_\th)}{\ln\rho}\right|_s
  = \frac{\gamma_\CR P_{\CR} + \gamma_\th P_{\th}}{P_\CR+P_\th} ,
\end{equation}
Subsequently the radius of the explosion is modified as it depends on the
compressibility of the post-shock gas in the interior of the blast wave.

In the case of a single polytropic fluid, the shock radius of the
blast wave evolves self similarly according to
\begin{equation}
  \label{eq:ST}
r_{\rmn{s}}(t) = \left(\frac{E_1}{\alpha \rho_1}\right)^{1/5} t^{2/5},
\end{equation}
where $t$ is the time since explosion and $\alpha$ is a self-similarity
parameter that depends on the effective adiabatic index, which itself is a
function of CR shock acceleration efficiency $\zeta$. To determine this
relation, we run a set of simulations, varying $\zeta\in[0,1]$ in steps of
0.05. In each simulation, we determine the average shock radii at different
times and obtain $\alpha(\zeta)$ via equation~(\ref{eq:ST}).

In Appendix~\ref{sec:ST_analytics}, we numerically solve the self-similar, spherically
symmetric conservation equations of mass, momentum and energy to determine the
behavior of $\alpha(\gamma)$. We find an analytical fit to the solution of the
form
\begin{equation}
  \label{eq:alpha1}
  \alpha(\gamma) \approx \dfrac{16}{75}
  \left[ \dfrac{\pi (3\gamma-1)}{(\gamma-1)(\gamma+1)^2} -\dfrac{3}{8}\right],  
\end{equation} 
which has an accuracy of approximately 0.8\%. Combining $\alpha(\zeta)$
(obtained with our simulations with CR acceleration and via
equation~\ref{eq:ST}) and $\alpha(\gamma)$ (equation~\ref{eq:alpha1}), we arrive
at an expression of the effective adiabatic index as a function shock
acceleration efficiency, $\gamma(\zeta)$, as shown in Fig.~\ref{fig4:mach}. In
particular, an efficiency of $\zeta = 0.5$ corresponds to an effective ratio of
specific heats of $\gamma \simeq 1.408$. As seen in Fig.~\ref{fig4:mach}, the
simulations do not reproduce the limit $\gamma\to4/3$ for $\zeta\to1$ due to
residual thermalization. Note that this case is purely academic and likely not realised in
Nature. Hence, we fit our simulation values for $\zeta\leq0.65$ with an equation
of the form
\begin{equation}
 \label{eq:alpha2}
 \gamma(\zeta) = a + \dfrac{b}{c + \zeta},
\end{equation} 
subject to the boundary condition of $\gamma= 5/3$ for $\zeta = 0$ and 4/3 for
$\zeta = 1$. This allows to express the parameters $b$ and $c$ solely as a
function of $a$.  The corresponding parameters satisfying these requirements are
shown in Table~\ref{tab:2}. Adopting the rational approximation of these fitting
parameters (Table~\ref{tab:2}) we obtain
\begin{equation}
\gamma(\zeta) = \dfrac{4}{3} \left( \dfrac{8 \zeta + 5}{9 \zeta + 4} \right).
\end{equation}
Combining this result with equation~(\ref{eq:alpha1}), we get a comprehensive formula
for the self-similarity parameter in equation~(\ref{eq:ST}) as a function of the
acceleration efficiency $\zeta$:
\begin{equation}
\label{eq:alpha_final}
\alpha({\zeta}) = \dfrac{2}{25}
\left[\dfrac{72 \pi (9 \zeta + 4)^2 (23 \zeta + 16)}{(5 \zeta + 8) (59 \zeta + 32)^2} - 1\right].
\end{equation}

\begin{figure*}
\centering
\includegraphics[scale=1.1]{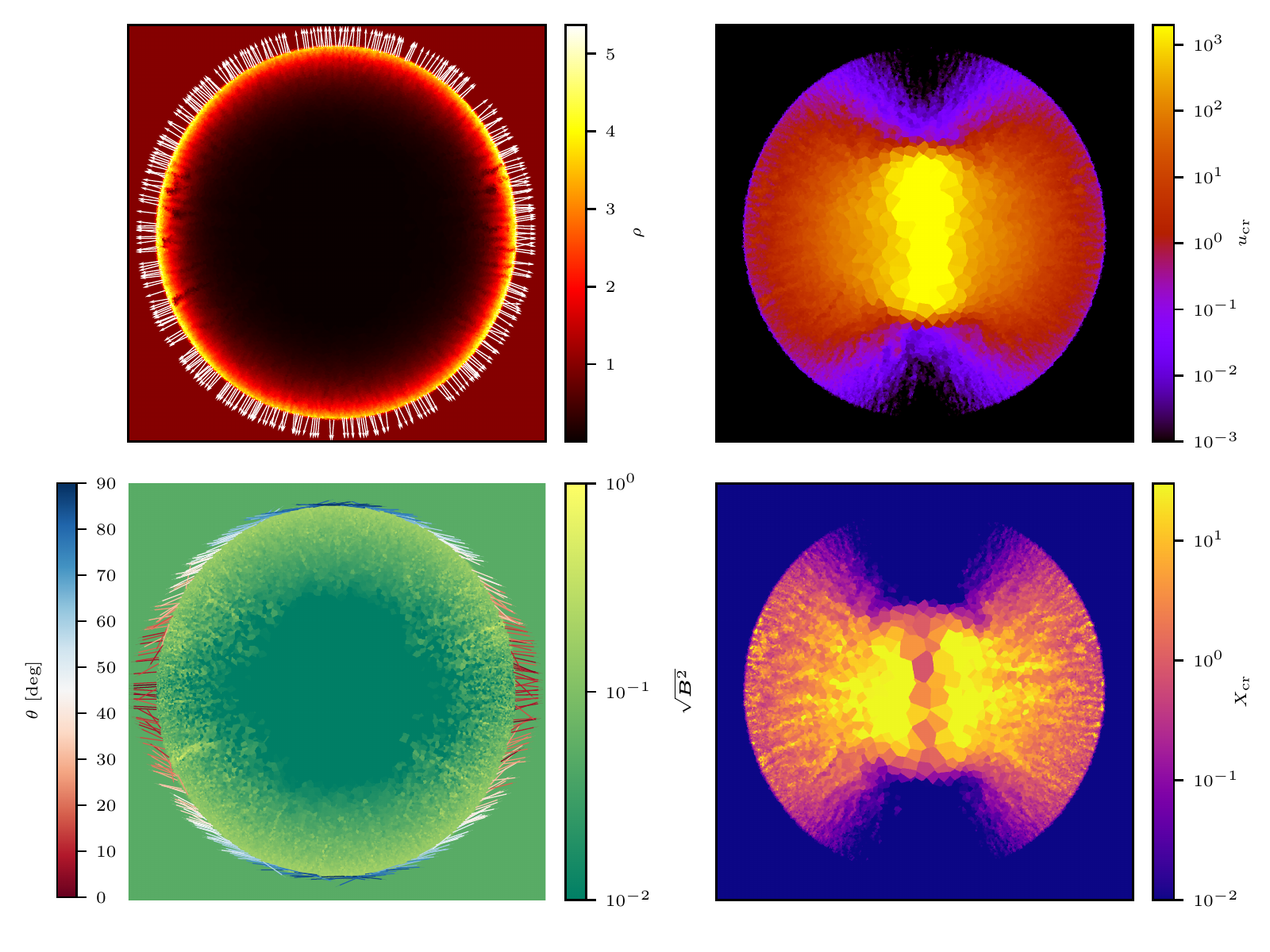} 
\caption{Sedov-Taylor blast wave with obliquity dependent CR acceleration
  expanding into a homogeneous magnetic field at $t=0.1$. Initially, the field
  is aligned with the horizontal direction. We show a 2D cross section of the
  density $\rho$ (top left) overplotted with normal vectors of the shock (as
  determined by our shock-finding algorithm); the specific CR energy
  $u_{\mathrm{cr}}$ (top right); the magnetic field strength
  $\sqrt{\vecbf{B}^2}$ (bottom left) overplotted with the outwards directed
  orientations of the magnetic field at the shocked cells (colour coded by
  magnetic obliquity $\theta$) and the CR-to-thermal pressure ratio
  $X_{\mathrm{cr}}$ (bottom right). We can see that the specific CR energy
  exhibits a quadrupolar anisotropy, with the maximum in the direction of the
  magnetic field (quasi-parallel shock configuration). The increased density in
  this quasi-parallel direction (due to the higher compressibility of the
  CR-enriched post-shock plasma) implies a slower shock expansion velocity and
  hence a slightly oblate shock surface with the two long axis aligned
  perpendicular to the ambient field direction. Note that the shock compression
  only amplifies the perpendicular magnetic field component, which re-aligns
  field vectors for oblique shocks.}
\label{fig5:sedov_maps}
\end{figure*}

\subsection{CR acceleration in a homogeneous field}
\label{sec:homogeneous}

Our initial Voronoi mesh is generated by randomly distributing mesh-generating
points in the unit box and relaxing the mesh via Lloyd's algorithm
\citeyearpar{journals/tit/Lloyd82}. The self-similarity of the problem, which is
not broken by CR acceleration \citep{2017MNRAS.465.4500P}, allows us to use
scale free units. We use a box of $200^3$ cells to ensure convergence also at
early times. Throughout the simulation box, we adopt a uniform density of
$\rho_1 = 1$, a negligible pressure of $P_1 = 10^{-4}$, a zero initial velocity,
and a thermal adiabatic index of $\gamma =5/3$.  At time $t=0$, we inject
thermal energy of $E_1=1$ into the central mesh cell.  We follow ideal MHD
without self-gravity and adopt a maximum acceleration efficiency of
$\zeta_0=0.5$ to amplify the (dynamical) effects of CR acceleration.

First, we adopt a homogeneous magnetic field in the box that is oriented along
the $x$ axis and a plasma beta of $\beta=1$. In Fig.~\ref{fig5:sedov_maps}
we show maps of different quantities in the equatorial plane at $t=0.1$, namely
the mass density $\rho$ (with the shock normal as measured in situ in the
simulations and shown in white), the specific CR energy $u_{\mathrm{cr}}$, the
magnetic field strength $\sqrt{\vecbf{B}^2}$ (with the magnetic orientations at
the shock colour coded by upstream magnetic obliquity), and the CR-to-thermal
pressure ratio $X_{\mathrm{cr}}=P_\CR/P_\th$.

\begin{figure*}
\centering
\includegraphics[scale=1.]{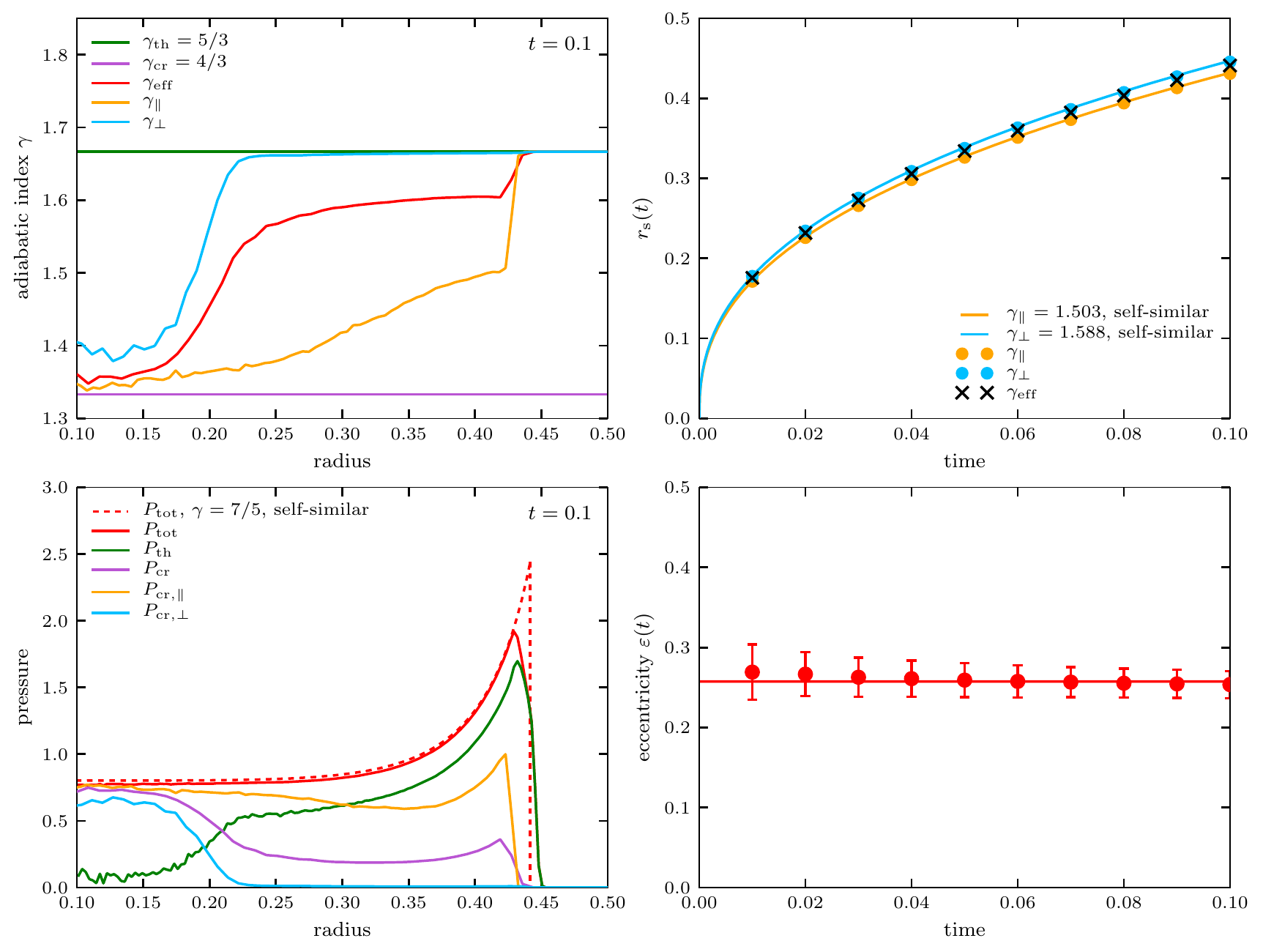} 
\caption{Radial profiles of characteristic quantities of the Sedov-Taylor
  explosion with obliquity dependent CR acceleration. The top left panel shows
  the radial profile of the effective adiabatic index. The adiabatic index in
  the direction perpendicular to the magnetic field (light blue) raises quickly
  to values comparable with the thermal adiabatic index $5/3$ (green) whereas
  the adiabatic index in the direction parallel to $\vecbf{B}$ (yellow) deviates
  only at larger radii from the relativistic value of $4/3$ (violet). The red
  line represents the effective adiabatic index averaged over all
  directions. The bottom left panel shows the corresponding radial profiles of
  the CR pressure in the different directions.  The thermal pressure dominates
  at larger radii while the CR pressure drops significantly outside the
  center. The top right panel represents the time evolution of the shock radius
  in the direction of the magnetic field (yellow) and perpendicular to it (light
  blue). Effective CR acceleration for quasi-parallel shock configurations
  yields an increased compressibility and hence a slower propagating shock. This
  is quantified in the time evolution of the shock eccentricity in the bottom
  right panel. The constant eccentricity of the oblate shock (within the
  uncertainties) demonstrates self-similar evolution of the blast wave also in
  this case.}
\label{fig6:sedov_line_maps}
\end{figure*}  

The unit vectors of the shock normal in the top left panel of
Fig.~\ref{fig5:sedov_maps} show a deviation from spherical symmetry with a
smaller shock radius and an enhanced density in the direction parallel and
anti-parallel to the magnetic field. This is the immediate consequence of
obliquity-dependent shock acceleration with copious CR production at
quasi-parallel shocks, which is accompanied by an increased compressibility due
to the softer equation of state of the composite fluid of CRs and thermal gas.
This is manifested in the quadrupolar morphology of $u_{\mathrm{cr}}$ with the axis
of symmetry aligned with the magnetic field orientation (top right of
Fig.~\ref{fig5:sedov_maps}). The morphology of $u_{\mathrm{cr}}$ is echoed by
$X_{\mathrm{cr}}$ (bottom right of Fig.~\ref{fig5:sedov_maps}). Adiabatic
expansion of a composite of CRs and thermal gas eventually yields a dominating
CR pressure in the interior of the explosion for quasi-parallel shock
geometries, at $|\theta|\lesssim\pi/4$.

An oblique shock only amplifies the perpendicular field component and leaves the
parallel component invariant. This re-orients the oblique magnetic field towards
the shock surface and increases the field strength at quasi-perpendicular shocks
(bottom left of Fig.~\ref{fig5:sedov_maps}). Our strongly magnetised background
plasma with $\beta=1$ becomes very weakly magnetised at the shock since the
adiabatic increase of magnetic pressure falls orders of magnitudes short in
comparison to the shock-dissipated thermal pressure at our strong Sedov-Taylor
shock. Hence, the magnetic field merely impacts the dynamics of the blast wave
through the magnetic obliquity-dependent shock acceleration of CRs and not
through its pressure.

This analysis is quantified in Fig.~\ref{fig6:sedov_line_maps}, where we show
radial profiles of different volume-weighted quantities, such as the effective
ratio of heat capacities $\gamma_{\rmn{eff}}$, the pressure, and the time evolution of the
shock radius and eccentricity of the oblate blast wave. The effective ratio of
heat capacities $\gamma_{\rmn{eff}}$ is computed from volume-averaged partial
pressures of the CR and thermal gas components via equation~\ref{eq:gamma_i}.
The top and bottom left panels of Fig.~\ref{fig6:sedov_line_maps} show the
radial variation of the effective adiabatic index $\gamma_{\parallel,\perp}$ and the partial
pressures $P_{\parallel,\perp}$ for two regions: parallel and perpendicular. $\gamma_\parallel$ and
$P_{\CR,\parallel}$ are computed from cells that belong to the hourglass-shaped
region inside the blast wave that was overrun by a quasi-parallel shock.  Here,
we define this quasi-parallel shocked region as a narrow double cone oriented
along the original magnetic field with an opening angle of
20$^\circ$. Similarly, we define the region overrun by quasi-perpendicular
shocks as the complement of a wide double cone that is bounded by an equatorial
band with latitude 20$^\circ$.
\begin{figure*}
\centering
\includegraphics[scale=0.95]{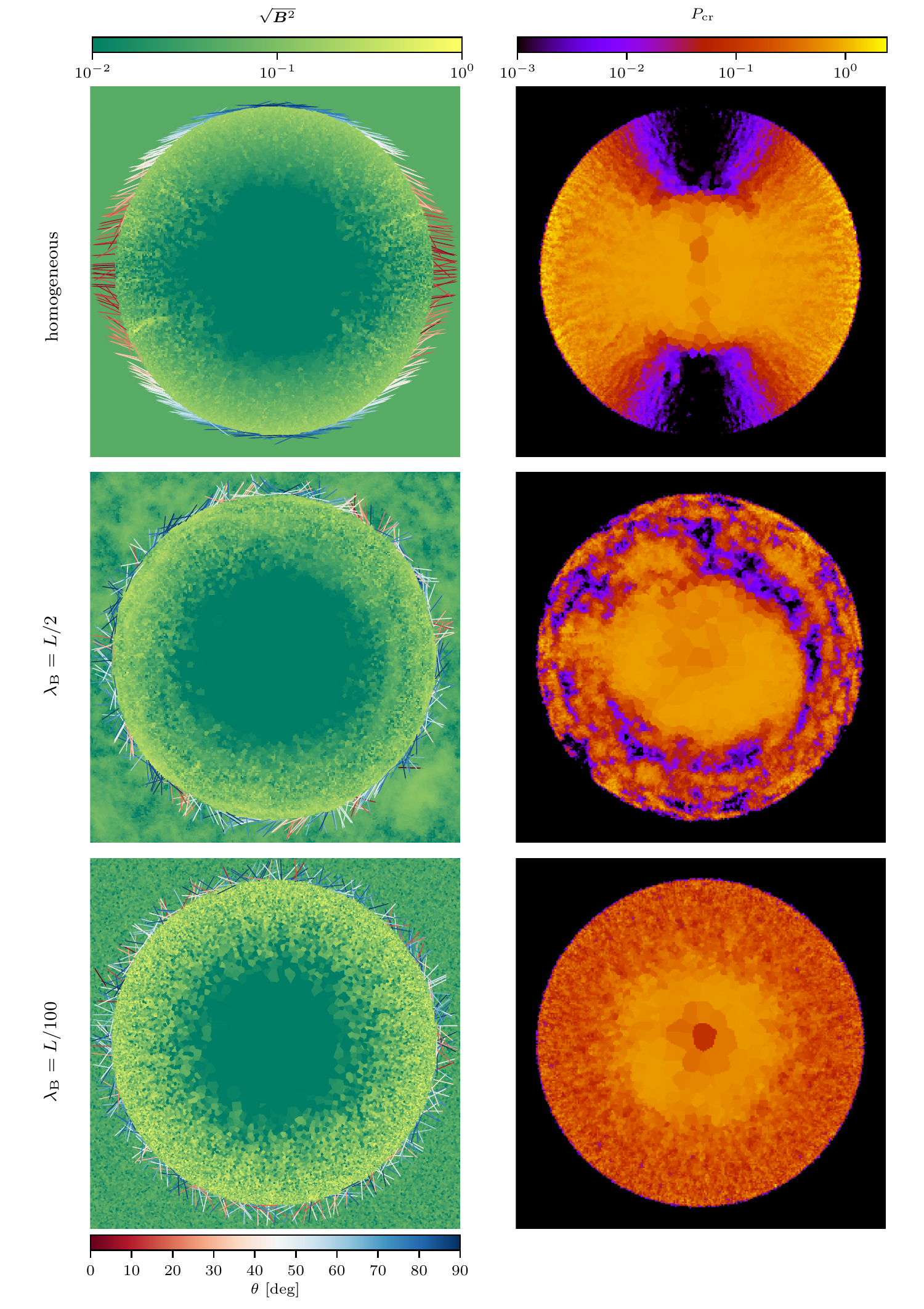} 
\caption{Cross-sections through the centre of 3D simulations of Sedov-Taylor
  explosions with obliquity-dependent CR acceleration. We show a simulation with
  a homogeneous, horizontally oriented magnetic field (first row), a turbulent
  magnetic field with a large correlation length of $\lambda_B=L/2$ (second row)
  and with a very small correlation length, $\lambda_B=L/100$, in comparison to
  the radius of the blast wave (third row). We depict magnetic field strength
  (left column) and CR pressure (right column). In the homogeneous field case we
  notice a quadrupolar CR distribution that is maximized for quasi-parallel
  shocks (visualised with red arrows in the left-hand panels) while the magnetic
  field is only adiabatically amplified at quasi-perpendicular shocks (blue
  arrows). In the second row we observe a patchy CR distribution with maxima at
  regions that were over-run with quasi-parallel shocks. The CR distribution in
  the case of small-scale turbulence (last row) is completely isotropic and the
  compression of the magnetic field is uniformly distributed across the shock.}
\label{fig7:6_plots_turbulent}
\end{figure*}
The copious CR production at a quasi-parallel shock with the subsequent
adiabatic expansion softens the adiabatic index to values close to that of a
fully relativistic gas of 4/3. Since the region overrun by a quasi-perpendicular
shock is characterized by a ratio of heat capacities close to a purely thermal
gas, the effective adiabatic index $\gamma_{\rmn{eff}}$ (shown in red) as well
as the spherically averaged CR pressure (shown in purple) levels off at values
in between.

The top-right panel of Fig.~\ref{fig6:sedov_line_maps} shows the time evolution
of the simulated shock radius (filled circles) and the self-similar analytic
solution (continuous lines). In line with the previous discussion, the shock
radius in the direction perpendicular to the ambient magnetic field moves faster
than the shock in the (anti-)parallel direction owing to the increased
compressibility of the latter due to efficient CR acceleration. The continuous
lines are obtained by fitting the shock radius evolution (equation~\ref{eq:ST})
in double-logarithmic space for $\alpha(\gamma)$.  Inverting
equation~\ref{eq:alpha1} yields the corresponding effective adiabatic factor
shown in the figure. In between those two curves, we show the solution for the
effective adiabatic index (crosses).

The increased compressibility of CR-enriched quasi-parallel shocks implies an
oblate shock surface that is characterised by an eccentricity, defined as
\begin{equation}
\epsilon(t) = \sqrt{1 - \left[\dfrac{r_\|(t)}{r_\bot (t)}\right]^2}.
\end{equation}
Due to the volumetric distribution of CRs with respect to the thermal gas, the
influence of CR production affects the entire explosion. This renders it
impossible to separate the cases of purely thermal and maximally efficient CR
acceleration for the perpendicular and parallel shock radii, respectively. This
means that a direct measurement of the eccentricity assuming a pure thermal
$\zeta = 0$ in the perpendicular direction and a CR-modulated $\zeta = \zeta_0$
in the parallel direction yields an incorrect result. Instead, an average
efficiency $\langle \zeta \rangle$ is required to determine the average radius
of the explosion, representing an intermediate case between the parallel and
perpendicular shock radius. In the lower right panel of
Fig.~\ref{fig6:sedov_line_maps}, we show the eccentricity of the oblate shock
surface along with the uncertainty intervals assuming Gaussian statistics,
\begin{equation}
  \Delta \epsilon = \left( \dfrac{1- \epsilon^2}{\epsilon}\right)
  \sqrt{\left( \dfrac{\sigma_\parallel}{r_\parallel}\right)^2 + 
    \left( \dfrac{\sigma_\perp}{r_\perp}\right)^2}
\end{equation}
where $\sigma_\parallel$ and $\sigma_\perp$ are the standard deviations of the
shock radius in the direction of the magnetic field and perpendicular to it,
respectively. We obtain the theoretical estimate for the eccentricity from our
measured self-similar solutions for the parallel and perpendicular shock radii
of the top-right panel in Fig.~\ref{fig6:sedov_line_maps}.  There are two
strategies to measure the eccentricity: first, determining the distance to the
cells of the shock surface inside narrow cones or bands of equal latitude that
are centered on the explosion and oriented along the magnetic field direction;
second: measuring the momenta of inertia of the entire oblate shock surface,
diagonalising the resulting tensor, determining the resulting eigenvalues and
extracting the length of the three semi-axes. We decided in favor of the first
method because it generates less numerical fluctuations.

We find a constant eccentricity of $\epsilon = 0.25 \pm 0.02$ during the adiabatic
expansion. Note that $\epsilon$ depends on the average efficiency
$\langle\zeta\rangle$ and is expected to be smaller for realistic maximum
acceleration efficiencies of order 0.15. The constant eccentricity with time
demonstrates that the Sedov-Taylor explosion remains self similar also in the
presence of obliquity-dependent CR acceleration.  In
Appendix~\ref{sec:convergence} we show that the measured eccentricity in our
simulations is numerically converged for $100^3$ grid cells.

\subsection{CR acceleration in a turbulent field}
\label{sec:turb}

After studying magnetic obliquity-dependent CR acceleration at a Sedov-Taylor
blast wave that propagates in a homogeneous magnetic field, we now turn to
turbulent magnetic fields with different magnetic correlation lengths
$\lambda_B=2\pi/k_{\rmn{inj}}$. As initial conditions for the magnetic field, we
adopt a Gaussian random field with a Kolmogorov power spectrum on scales smaller
than the coherence length and a white-noise power spectrum on larger scales, as
described in Sect.~\ref{sec:turb_ICs}.  The larger $\lambda_B$ in comparison to
the shock radius, the fewer statistically independent regions of correlated
magnetic fields there are inside the blast wave. Hence we introduce the magnetic
coherence length in units of the shock radius,
$\Lambda_B=\lambda_B/r_{\rmn{s}}(t)$ as a new parameter. Blast waves with
the same $\Lambda_B$ are statistically self similar.

We perform several simulations with different correlation lengths ranging from
$\lambda_B = L$ to $\lambda_B = L/200$ for $200^3$-cell runs.  In
Fig.~\ref{fig7:6_plots_turbulent} we show different realizations of the CR
pressure for varying the correlation lengths of the magnetic field and compare
the results to our previous simulation with a homogeneous field.  Correlated
magnetic patches imply a similarly patchy CR distribution: regions that are
overrun by quasi-parallel shocks are CR enriched whereas regions that have
experienced quasi-perpendicular shocks result in voids without CRs.  As the
scaled correlation length $\Lambda_B$ becomes smaller, the magnetic obliquity
also changes on smaller scales and the number of CR islands becomes more
frequent to the point that they merge into a single (noisy) CR distribution.
This case is similar to uniform CR acceleration (which is independent of
magnetic obliquity), but exhibits a lower overall CR acceleration efficiency.

\subsection{Average CR acceleration efficiency and field realignment}

In order to understand the blast wave-averaged CR acceleration efficiency, we
consider the two limiting cases of a homogeneous field and a fully turbulent
field with a coherence scale of the initial grid resolution ($\lambda_B = \Delta
L$) analytically. In the small-scale turbulent case, we consider a fixed shock
normal pointing along $\bm{\hat{z}}$ without loss of generality. The magnetic
field vector can then assume any direction on the upper half-sphere because CR
acceleration does not depend on the sign of the magnetic field and is symmetric
with respect to $\theta=0$. Hence, the probability distribution of the magnetic
obliquity is given by $f(\theta)=\sin\theta$ with
$\theta\in[0,\pi/2]$. 

\begin{figure*}
\centering
\includegraphics[scale=0.7]{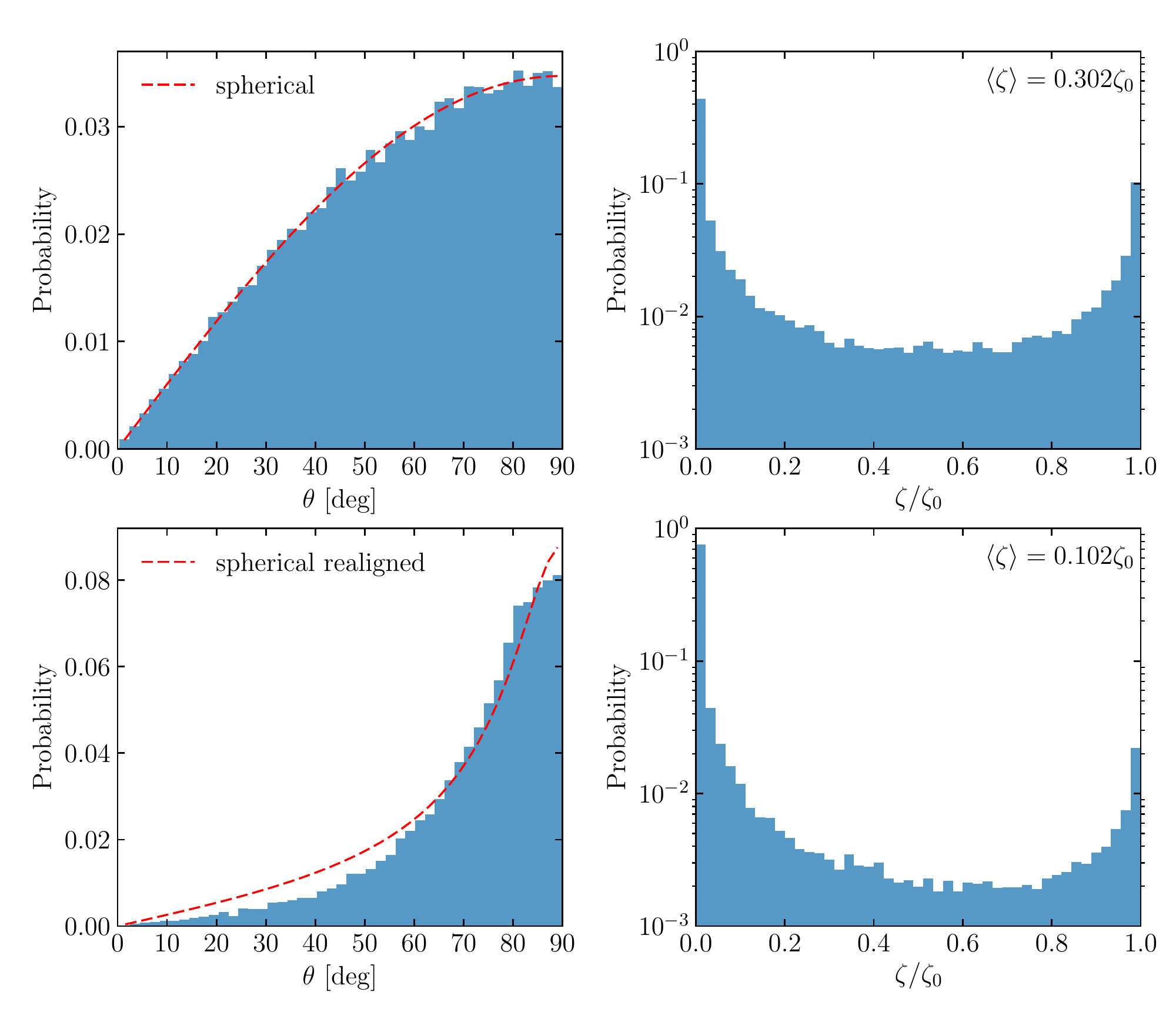} 
\caption{Probability distribution functions (PDFs) of the magnetic obliquity
  (left) and CR acceleration efficiency $\zeta$ (right) in the case of a
  homogeneous magnetic field in the upstream regime (top panels) and after
  accounting for magnetic re-orientation in the immediate downstream regime
  (bottom panels). Here, we adopt an artificially small value for the maximum CR
  acceleration efficiency of $\zeta_0 = 0.02$, which implies an almost spherical
  shock due to the negligible CR backreaction in this case.  The obliquity
  distribution follows the theoretical expectation of
  $f(\theta)=\sin(\theta)$. Accounting for magnetic re-orientation at oblique
  shocks skews this distribution towards quasi-perpendicular geometries (bottom
  left), which can be analytically described by equation~\eqref{eq:f(t(r))} for
  $\epsilon=0$ (dashed line). Note that $\zeta$ follows a bimodal distribution
  as a result of the flatness of the efficiency function
  (equation~\ref{eq:efficiency}) at quasi-perpendicular and -parallel shocks
  with a sharp transition in between.}
\label{fig8:histogram_l}
\end{figure*}

Integrating the efficiency over this probability
distribution results in the average efficiency according to
\begin{equation}
  \label{eq:average}
  \langle\zeta\rangle = \int_0^{\pi/2} \zeta (\theta) \sin \theta\, \de \theta =
  \begin{cases}
    \zeta_0\left(1-\frac{\displaystyle 1}{\displaystyle \sqrt{2}}\right),
      & \mbox{for }\zeta= \zeta_{\rmn{toy}},\\
    0.302 \,\zeta_0, & \mbox{with Equ.~(\ref{eq:efficiency}).}
  \end{cases}
\end{equation}
Here, we introduced a toy example for the obliquity dependent
acceleration that is represented by a discontinuous jump of the efficiency at
$\theta_{\rmn{crit}}$ from $\zeta_0$ to zero:
\begin{equation}
\zeta_{\rmn{toy}}(\theta)=\zeta_0\Theta(\theta_{\rmn{crit}}-\theta),
\end{equation}
where $\Theta(x)$ is the Heaviside function, representing the limiting case of
$\delta \rightarrow \infty$ in equation~(\ref{eq:efficiency}). This gives us a
lower limit for the efficiency.

In the case of a homogeneous field, we fix the magnetic field vector in space
and point it into the $z$ direction without loss of generality. Again, the shock
normal can assume any direction on the upper half-sphere so that we obtain the
same probability distribution function as in the small-scale turbulent case,
$f(\theta)=\sin\theta$. The average CR shock acceleration efficiency is thus
also given by equation~(\ref{eq:average}).

We find that eccentricity plays an important role in shaping the probability
distribution of the obliquity. To take this into account we define an
ellipsoidal reference frame via
\begin{equation}
x =  \left(h  + \dfrac{a}{\sqrt{1 - \epsilon^2 \cos^2 \varphi}}\right) \sin\varphi \cos\lambda,
\end{equation}
\begin{equation}
y =  \left(h  + \dfrac{a}{\sqrt{1 - \epsilon^2 \cos^2 \varphi}}\right) \sin\varphi \sin\lambda,
\end{equation}
\begin{equation}
z = \left[ h + \dfrac{(1-\epsilon^2)a}{\sqrt{1-\epsilon^2 \cos^2 \varphi}} \right] \cos\varphi,
\end{equation}

\begin{figure*}
\centering
\includegraphics[scale=0.7]{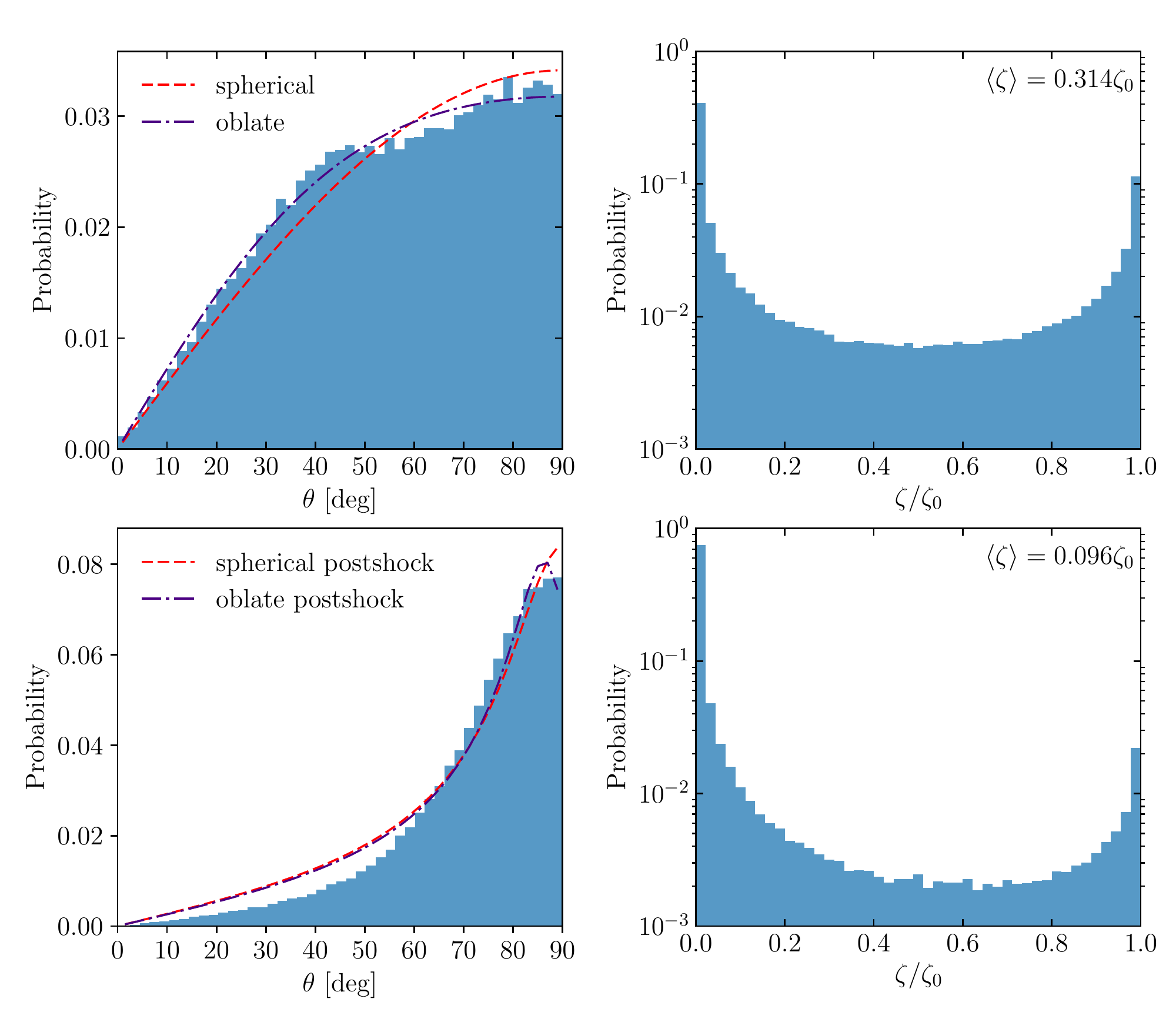} 
\caption{PDF of the magnetic obliquity (left) and CR acceleration efficiency
  $\zeta$ (right) in the case of a homogeneous magnetic field in the upstream
  regime (top panels) and after accounting for magnetic re-orientation (bottom
  panels) with a maximum efficiency of $\zeta_0 = 0.5$.  The obliquity
  distribution in the top left panel follows our theoretical prediction
  $f(\theta,\epsilon)$ (purple, equation~\ref{eq:ell_sin_norm}) of an oblate
  expanding shock. For comparison, we also show the spherical PDF (red dashed,
  see Fig.~\ref{fig8:histogram_l}). Accounting for magnetic re-orientation at
  oblique shocks skews this distribution towards quasi-perpendicular geometries
  (bottom left), which can be analytically described by
  equation~\eqref{eq:f(t(r))}. }
\label{fig9:histogram_h}
\end{figure*}

where $\lambda$ and $\varphi$ are the longitude and pseudo-latitude from the
ellipsoid, respectively, $h$ is the height above the surface of the ellipsoid,
$a$ the semi-major axis, and $\epsilon$ the eccentricity.  

As shown in Appendix~\ref{sec:ellipsoid}, for a homogeneous magnetic field that points into the
positive $z$ direction (short axis of the oblate ellipsoid) the angle $\varphi$
is by construction equal to the definition of the magnetic obliquity $\theta$.
Using the fact that $\varphi \equiv \theta$, the Jacobian of this coordinate
transformation on the oblate surface ($h=0$) is given by
\begin{equation}
\label{eq:ell_sin}
S(\theta,\epsilon) =  \sin \theta
 \dfrac{1-\epsilon^2}{(1 - \epsilon^2 \cos^2\theta)^2}.
\end{equation}
Hence, the normalized distribution function for the obliquity $\theta$ reads
\begin{equation}
\label{eq:ell_sin_norm}
f(\theta,\epsilon) =  S(\theta,\epsilon)
\left[  \int_0^{\pi/2} S(\theta,\epsilon) \de \theta \right]^{-1},
\end{equation}
which reduces to $\sin\theta$ for $\epsilon = 0$. 

For our simulations with
$\zeta_0=0.5$, we obtain an eccentricity of $\tilde{\epsilon} = 0.25$, and hence an
average efficiency of
\begin{equation}
\label{eq:elliptic_eff}
\langle \zeta (\tilde{\epsilon})\rangle = \int_0^{\pi/2} \zeta(\theta) f(\theta,\tilde{\epsilon}) \de \theta= 0.316 \zeta_0.
\end{equation}
The error $\Delta \zeta $ on this quantity derives from the uncertainty on the eccentricity:
\begin{equation}
\Delta \tilde{\zeta} = \left.\dfrac{\di \langle \zeta(\epsilon)\rangle}{\di \epsilon}\right|_{\tilde{\epsilon}} \Delta\tilde{\epsilon} =  \left.\left[\int_0^\pi \zeta(\theta) \dfrac{\di f(\theta,\epsilon)}{\di \epsilon} \right|_{\tilde{\epsilon}} \de \theta \right] \Delta \tilde{\epsilon} = 0.002 \zeta_0 ,
\end{equation}
such that the efficiency for this oblate reads
\begin{equation}
\tilde{\zeta} = (0.316 \pm 0.002) \zeta_0 .
\end{equation}
Because we propagate the upstream value of the magnetic obliquity to the shock
surface, the resulting obliquity distribution at the shock is expected to follow
$f(\theta)=\sin(\theta)$ in the case of a sphere and
equation~\eqref{eq:ell_sin_norm} for ellipsoids.

\begin{figure}
\centering
\includegraphics[scale=0.7]{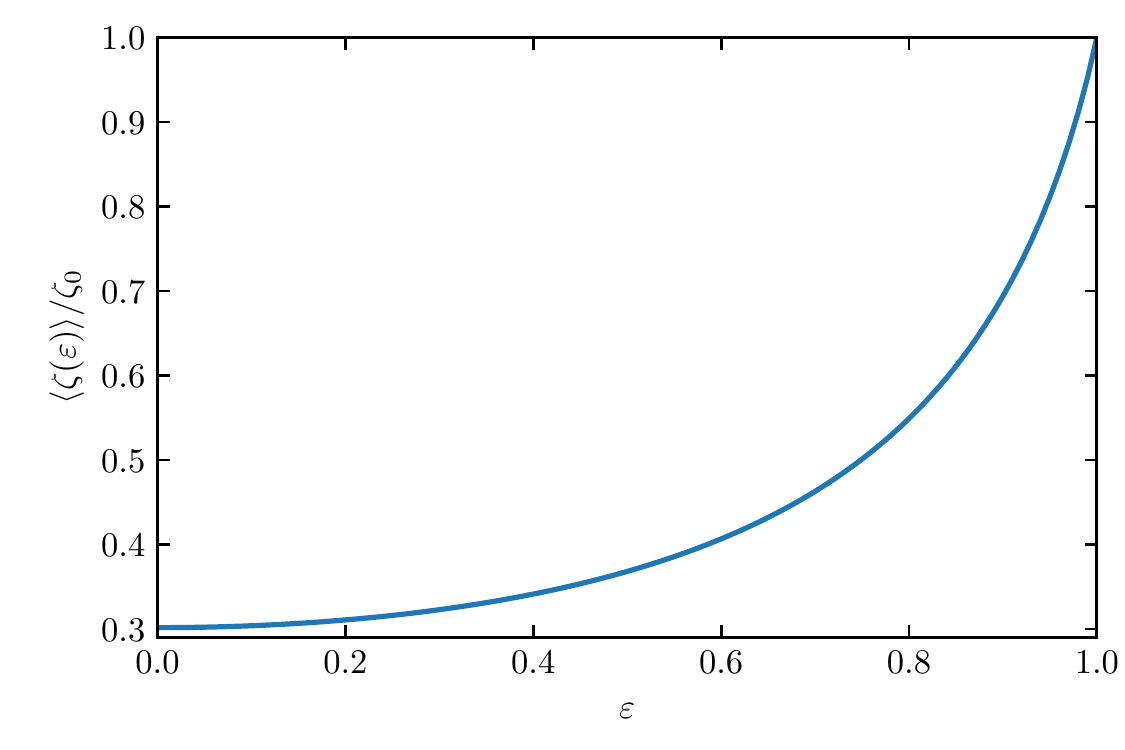} 
\caption{Average acceleration efficiency as a function of the eccentricity
  $\epsilon$ of the oblate according to the distribution shown in
  equation~(\ref{eq:ell_sin}) and equation~(\ref{eq:elliptic_eff}) in a homogeneous
  magnetic field configuration. For $\epsilon=0$ the original value of
  equation~\eqref{eq:average} is restored while for an eccentricity of unity the
  oblate degenerates into a circle, yielding everywhere a maximum efficiency.}
\label{fig10:ell}
\end{figure}

\begin{figure}
\centering
\includegraphics[scale=0.8]{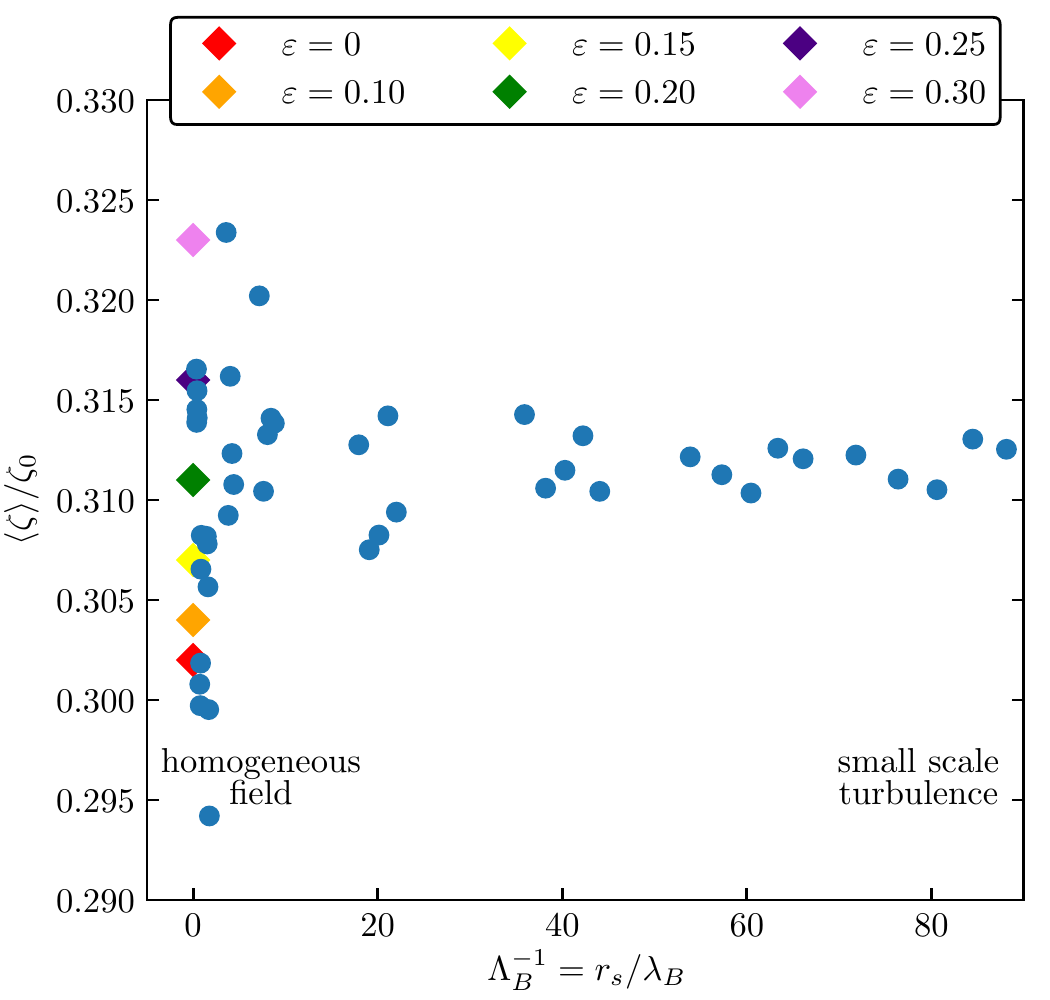} 
\caption{Average CR acceleration efficiency for a turbulent magnetic field with
  different correlation lengths $\Lambda_B=\lambda_B/r_{\rmn{s}}(t)$.  The case
  $\Lambda_B^{-1}=0$ corresponds to a homogeneous magnetic field and
  $\Lambda_B^{-1}=85$ relates to the small-scale turbulent case with a coherence
  scale equal to the initial grid resolution of our simulation. The coloured
  diamonds at $\Lambda_B^{-1} = 0$ show the homogeneous field case for different
  average CR acceleration efficiencies corresponding to different
  eccentricities. The simulation points scatter around the value $\langle \zeta
  \rangle = (0.311 \pm 0.005)\zeta_0$.}
  \label{fig11:last}
\end{figure}

In the next step, we assess the distribution of downstream magnetic obliquity as
a result of realignment of the tangential component of the magnetic field due to
the shock.  We can estimate the obliquity after magnetic realignment with the
aid of the Rankine-Hugoniot jump conditions:
\begin{equation}
\label{eq:mag_rel}
\cos\theta_2 (r_{\rmn{c}})=
\dfrac{\cos\theta_1}{\sqrt{\cos^2\theta_1 + r_{\rmn{c}}^{-2} \sin^2\theta_1}}
\end{equation}
where $r_{\rmn{c}}$ is the compression ratio at the shock. We can then insert
this formula into equation~\eqref{eq:ell_sin} to obtain the expected distribution
of realigned angles for oblates:
\begin{equation}
\label{eq:f(t(r))}
f[\theta(r_{\rmn{c}}),\epsilon] = S[\theta(r_{\rmn{c}}),\epsilon]
\left[ \int_0^{\pi/2} S[\theta(r_{\rmn{c}}),\epsilon] \de \theta \right]^{-1}
\end{equation}
which reduces to $\sin[\theta(r_{\rmn{c}})]$ for $\epsilon = 0$. We find good
agreement of these theoretical distributions and our simulations in
Figs.~\ref{fig8:histogram_l} and \ref{fig9:histogram_h} for a maximum efficiency
of $\zeta_0 = 0.02$ and of $\zeta_0 = 0.5$, respectively.  Note that the low CR
acceleration efficiency in Fig.~\ref{fig8:histogram_l} allows to neglect the
geometrical anisotropy of the shock surface that results from copious CR
production in the direction of the magnetic field. In all cases, we find a
bimodal distribution of $\zeta$ as a result of the flat efficiencies at
quasi-perpendicular and -parallel shocks with a sharp transition in between.  We
also find a good agreement of the realigned obliquity distributions, which are
skewed towards quasi-perpendicular geometries.

The effect of an oblate geometry becomes evident in Fig.~\ref{fig9:histogram_h}
(upper left panel), which shows an improved fit of the elliptical distribution
in comparison to the spherical sinusoidal distribution. However, the resulting
shape of the efficiency distribution is only little affected. Thus, its average
value is only slightly increased over the spherical case as can be inferred from
Figs.~\ref{fig8:histogram_l} and \ref{fig9:histogram_h}. The full evolution of
$\langle \zeta \rangle$ as a function of $\epsilon$ is shown in
Fig.~\ref{fig10:ell}. The academic case of $\epsilon=1$ represents an unphysical
limit where the oblate degenerates into a circle. This orients the shock surface
in such a way so that it is always parallel to the direction of the magnetic
field and yields the maximum possible acceleration efficiency.

In Fig.~\ref{fig11:last} we summarize results for different simulations with
varying correlation lengths of $\lambda_B^{-1} = [2,4,10,20,50,100,150,200]$
each at times $t=[5,6,7,8,9,10]\times0.01$ in simulation units.  The expanding
shock front starts to embrace more and more coherent magnetic patches whose
number scales as $N \propto \Lambda_B^{-3}$. For comparison, we also show the
theoretically expected eccentricities for the homogeneous field case with
coloured diamonds. There is no trend in the evolution of the average
acceleration efficiency as a function of $\Lambda_B^{-1}$. Instead, the
simulation values scatter between an eccentricity of $\epsilon = 0.10$ (orange
diamond) and $\epsilon=0.3$ (violet diamond), yielding a value of $\langle \zeta
\rangle = (0.311 \pm 0.05) \zeta_0$.  For $\lambda_B\gg\Delta L$, CR-rich
patches give rise to corrugations of the shock surface causing local small
deviations from spherical symmetry. If the correlation length becomes comparable
to the initial grid resolution then the blast wave becomes spherical, albeit
with a slightly higher efficiency.

\section{Conclusions}
\label{sec:conclusions}

In this paper we perform MHD simulations of the evolution of supernova remnants
in the Sedov-Taylor phase. For the first time, we model magnetic obliquity
dependent CR acceleration and study i) its dynamical effects on the overall
evolution of the blast wave and ii) how different magnetic geometries affect the
resulting CR distribution. To this end, we use results from hybrid PIC
simulations (with kinetic ions and fluid electrons) of non-relativistic, large
Mach number shocks. Those demonstrate that only quasi-parallel magnetic shock
configurations can accelerate ions while quasi-perpendicular shocks are
ineffective.

Using idealized shock tube experiments, we show that our algorithm is able to
recover the input direction of the magnetic field with a Gaussian scatter of
around $3^\circ$. When we change the magnetic orientation from
quasi-perpendicular to quasi-parallel configurations, the efficiency of CR
acceleration and the associated post-shock compressibility increase. This leads
to density jumps that exceed the theoretical limit $\rho_2/\rho_1=4$ (valid for
a thermal gas), slows down the shock and decreases the Mach number.

We derive analytical exact solutions of the Sedov-Taylor blast-wave problem with
CR acceleration (neglecting obliquity dependent effects). We numerically solve
the self-similar, spherically symmetric conservation equations of mass, momentum
and energy to determine the behavior of the shock radius. This enables us to
derive analytical fitting functions for the effective ratio of specific heats
for a composite of thermal gas and CRs as a function of the maximum acceleration
efficiency.

Our simulations of the Sedov-Taylor blast wave problem with obliquity dependent
CR acceleration in a homogeneous magnetic field geometry show the emergence of
an oblate ellipsoidal shock surface. Its short axis is aligned with the ambient
magnetic field orientation due to the efficient CR acceleration at
quasi-parallel shocks.  The ellipsoidal shock surface has an eccentricity of
$\epsilon = 0.25 \pm 0.02$ for a maximum CR acceleration efficiency of
$\zeta_0=0.5$ (which decreases for more realistic maximum efficiencies). The
shock eccentricity does not change with time, demonstrating that the
Sedov-Taylor explosion also remains self similar in the presence of
obliquity-dependent CR acceleration.  Because an oblique shock only amplifies
the perpendicular field component, this re-orients an oblique magnetic field
towards the shock surface. We find that this re-orientation effect has no
practical influence on the average CR acceleration efficiency because the
acceleration efficiency exhibits two flat plateaus at quasi-parallel and
-perpendicular shocks and a fast transition in between.

Sedov-Taylor explosions in a turbulent magnetic field yield a patchy CR
distribution with tangential, filamentary overdensities delineating regions that
were over-run by quasi-parallel shocks and filamentary patches devoid of CRs,
which were swept by quasi-perpendicular shocks. The CR distribution becomes
completely isotropic if the magnetic turbulence exhibits a very small coherence
scale in comparison to the shock radius. We derive the averaged CR acceleration
efficiency to $\approx0.3$ of the maximum CR acceleration efficiency for our
adopted CR efficiency function, independent of coherence scale.

In particular, the peculiar morphology of the CR pressure distribution that
result from obliquity-dependent CR acceleration in a turbulent magnetic field
could be the origin of the observed tangential filamentary morphology of some
shell-type middle-aged supernova remnants at TeV gamma rays. We will study this
effects in a separate publication.  We finally note that the fluctuating TeV
gamma-ray morphology would be a direct consequence of the obliquity dependent
acceleration in this picture and does not require large upstream CR
fluctuations or strong gradients in the ambient density; thereby opening the
possibility of realistically modelling supernova remnants at gamma rays in the future.

\section{Acknowledgments}
It is a pleasure to thank Kevin Schaal for his help on the numerics.  We warmly
thank V. Springel for the use of \AREPO.  This work has been supported by the
European Research Council under ERC-CoG grant CRAGSMAN-646955.

\bibliographystyle{mnras}
\bibliography{ms}

\begin{thebibliography}{}
\makeatletter
\relax
\def\mn@urlcharsother{\let\do\@makeother \do\$\do\&\do\#\do\^\do\_\do\%\do\~}
\def\mn@doi{\begingroup\mn@urlcharsother \@ifnextchar [ {\mn@doi@}
  {\mn@doi@[]}}
\def\mn@doi@[#1]#2{\def\@tempa{#1}\ifx\@tempa\@empty \href
  {http://dx.doi.org/#2} {doi:#2}\else \href {http://dx.doi.org/#2} {#1}\fi
  \endgroup}
\def\mn@eprint#1#2{\mn@eprint@#1:#2::\@nil}
\def\mn@eprint@arXiv#1{\href {http://arxiv.org/abs/#1} {{\tt arXiv:#1}}}
\def\mn@eprint@dblp#1{\href {http://dblp.uni-trier.de/rec/bibtex/#1.xml}
  {dblp:#1}}
\def\mn@eprint@#1:#2:#3:#4\@nil{\def\@tempa {#1}\def\@tempb {#2}\def\@tempc
  {#3}\ifx \@tempc \@empty \let \@tempc \@tempb \let \@tempb \@tempa \fi \ifx
  \@tempb \@empty \def\@tempb {arXiv}\fi \@ifundefined
  {mn@eprint@\@tempb}{\@tempb:\@tempc}{\expandafter \expandafter \csname
  mn@eprint@\@tempb\endcsname \expandafter{\@tempc}}}

\bibitem[\protect\citeauthoryear{{Axford}, {Leer}  \& {Skadron}}{{Axford}
  et~al.}{1977}]{1977ICRC...11..132A}
{Axford} W.~I.,  {Leer} E.,   {Skadron} G.,  1977, International Cosmic Ray
  Conference, \href {http://adsabs.harvard.edu/abs/1977ICRC...11..132A} {11,
  132}

\bibitem[\protect\citeauthoryear{{Bell}}{{Bell}}{1978}]{1978MNRAS.182..147B}
{Bell} A.~R.,  1978, \mn@doi [\mnras] {10.1093/mnras/182.2.147}, \href
  {http://adsabs.harvard.edu/abs/1978MNRAS.182..147B} {182, 147}

\bibitem[\protect\citeauthoryear{{Bell}}{{Bell}}{2004}]{2004MNRAS.353..550B}
{Bell} A.~R.,  2004, \mn@doi [\mnras] {10.1111/j.1365-2966.2004.08097.x}, \href
  {http://adsabs.harvard.edu/abs/2004MNRAS.353..550B} {353, 550}

\bibitem[\protect\citeauthoryear{{Bell}}{{Bell}}{2015}]{2015MNRAS.447.2224B}
{Bell} A.~R.,  2015, \mn@doi [\mnras] {10.1093/mnras/stu2596}, \href
  {http://adsabs.harvard.edu/abs/2015MNRAS.447.2224B} {447, 2224}

\bibitem[\protect\citeauthoryear{{Bell}, {Schure}, {Reville}  \&
  {Giacinti}}{{Bell} et~al.}{2013}]{2013MNRAS.431..415B}
{Bell} A.~R.,  {Schure} K.~M.,  {Reville} B.,   {Giacinti} G.,  2013, \mn@doi
  [\mnras] {10.1093/mnras/stt179}, \href
  {http://adsabs.harvard.edu/abs/2013MNRAS.431..415B} {431, 415}

\bibitem[\protect\citeauthoryear{{Beshley} \& {Petruk}}{{Beshley} \&
  {Petruk}}{2012}]{2012MNRAS.419.1421B}
{Beshley} V.,  {Petruk} O.,  2012, \mn@doi [\mnras]
  {10.1111/j.1365-2966.2011.19799.x}, \href
  {http://adsabs.harvard.edu/abs/2012MNRAS.419.1421B} {419, 1421}

\bibitem[\protect\citeauthoryear{{Blandford} \& {Ostriker}}{{Blandford} \&
  {Ostriker}}{1978}]{1978ApJ...221L..29B}
{Blandford} R.~D.,  {Ostriker} J.~P.,  1978, \mn@doi [\apjl] {10.1086/182658},
  \href {http://adsabs.harvard.edu/abs/1978ApJ...221L..29B} {221, L29}

\bibitem[\protect\citeauthoryear{{Bohdan}, {Niemiec}, {Kobzar}  \&
  {Pohl}}{{Bohdan} et~al.}{2017}]{2017ApJ...847...71B}
{Bohdan} A.,  {Niemiec} J.,  {Kobzar} O.,   {Pohl} M.,  2017, \mn@doi [\apj]
  {10.3847/1538-4357/aa872a}, \href
  {http://adsabs.harvard.edu/abs/2017ApJ...847...71B} {847, 71}

\bibitem[\protect\citeauthoryear{{Caprioli} \& {Spitkovsky}}{{Caprioli} \&
  {Spitkovsky}}{2014}]{2014ApJ...783...91C}
{Caprioli} D.,  {Spitkovsky} A.,  2014, \mn@doi [\apj]
  {10.1088/0004-637X/783/2/91}, \href
  {http://adsabs.harvard.edu/abs/2014ApJ...783...91C} {783, 91}

\bibitem[\protect\citeauthoryear{{Caprioli}, {Pop}  \& {Spitkovsky}}{{Caprioli}
  et~al.}{2015}]{2015ApJ...798L..28C}
{Caprioli} D.,  {Pop} A.-R.,   {Spitkovsky} A.,  2015, \mn@doi [\apjl]
  {10.1088/2041-8205/798/2/L28}, \href
  {http://adsabs.harvard.edu/abs/2015ApJ...798L..28C} {798, L28}

\bibitem[\protect\citeauthoryear{{Castro}, {Slane}, {Patnaude}  \&
  {Ellison}}{{Castro} et~al.}{2011}]{2011ApJ...734...85C}
{Castro} D.,  {Slane} P.,  {Patnaude} D.~J.,   {Ellison} D.~C.,  2011, \mn@doi
  [\apj] {10.1088/0004-637X/734/2/85}, \href
  {http://adsabs.harvard.edu/abs/2011ApJ...734...85C} {734, 85}

\bibitem[\protect\citeauthoryear{{Chevalier}}{{Chevalier}}{1983}]{1983ApJ...272..765C}
{Chevalier} R.~A.,  1983, \mn@doi [\apj] {10.1086/161338}, \href
  {http://adsabs.harvard.edu/abs/1983ApJ...272..765C} {272, 765}

\bibitem[\protect\citeauthoryear{{Guo}, {Sironi}  \& {Narayan}}{{Guo}
  et~al.}{2014a}]{2014ApJ...794..153G}
{Guo} X.,  {Sironi} L.,   {Narayan} R.,  2014a, \mn@doi [\apj]
  {10.1088/0004-637X/794/2/153}, \href
  {http://adsabs.harvard.edu/abs/2014ApJ...794..153G} {794, 153}

\bibitem[\protect\citeauthoryear{{Guo}, {Sironi}  \& {Narayan}}{{Guo}
  et~al.}{2014b}]{2014ApJ...797...47G}
{Guo} X.,  {Sironi} L.,   {Narayan} R.,  2014b, \mn@doi [\apj]
  {10.1088/0004-637X/797/1/47}, \href
  {http://adsabs.harvard.edu/abs/2014ApJ...797...47G} {797, 47}

\bibitem[\protect\citeauthoryear{{HESS Collaboration} et~al.,}{{HESS
  Collaboration} et~al.}{2016}]{2016Natur.531..476H}
{HESS Collaboration} et~al., 2016, \mn@doi [\nat] {10.1038/nature17147}, \href
  {http://adsabs.harvard.edu/abs/2016Natur.531..476H} {531, 476}

\bibitem[\protect\citeauthoryear{{Kang} \& {Ryu}}{{Kang} \&
  {Ryu}}{2013}]{2013ApJ...764...95K}
{Kang} H.,  {Ryu} D.,  2013, \mn@doi [\apj] {10.1088/0004-637X/764/1/95}, \href
  {http://adsabs.harvard.edu/abs/2013ApJ...764...95K} {764, 95}

\bibitem[\protect\citeauthoryear{{Krymskii}}{{Krymskii}}{1977}]{1977DoSSR.234.1306K}
{Krymskii} G.~F.,  1977, Akademiia Nauk SSSR Doklady, \href
  {http://adsabs.harvard.edu/abs/1977DoSSR.234.1306K} {234, 1306}

\bibitem[\protect\citeauthoryear{{Landau} \& {Lifshitz}}{{Landau} \&
  {Lifshitz}}{1966}]{1966hydr.book.....L}
{Landau} L.~D.,  {Lifshitz} E.~M.,  1966, {Hydrodynamik}

\bibitem[\protect\citeauthoryear{Lloyd}{Lloyd}{1982}]{journals/tit/Lloyd82}
Lloyd S.~P.,  1982, IEEE Trans. Information Theory, 28, 129

\bibitem[\protect\citeauthoryear{{Marcowith} et~al.,}{{Marcowith}
  et~al.}{2016}]{2016RPPh...79d6901M}
{Marcowith} A.,  et~al., 2016, \mn@doi [Reports on Progress in Physics]
  {10.1088/0034-4885/79/4/046901}, \href
  {http://adsabs.harvard.edu/abs/2016RPPh...79d6901M} {79, 046901}

\bibitem[\protect\citeauthoryear{{Mihalas} \& {Mihalas}}{{Mihalas} \&
  {Mihalas}}{1984}]{1984oup..book.....M}
{Mihalas} D.,  {Mihalas} B.~W.,  1984, {Foundations of radiation hydrodynamics}

\bibitem[\protect\citeauthoryear{{Miyoshi} \& {Kusano}}{{Miyoshi} \&
  {Kusano}}{2005}]{2005JCoPh.208..315M}
{Miyoshi} T.,  {Kusano} K.,  2005, \mn@doi [J. Comput. Phys.]
  {10.1016/j.jcp.2005.02.017}, \href
  {http://adsabs.harvard.edu/abs/2005JCoPh.208..315M} {208, 315}

\bibitem[\protect\citeauthoryear{{Neronov}}{{Neronov}}{2017}]{Neronov2017}
{Neronov} A.,  2017, \prl, \href
  {http://adsabs.harvard.edu/abs/2017arXiv171102734N} {119}

\bibitem[\protect\citeauthoryear{{Pakmor} \& {Springel}}{{Pakmor} \&
  {Springel}}{2013}]{2013MNRAS.432..176P}
{Pakmor} R.,  {Springel} V.,  2013, \mn@doi [\mnras] {10.1093/mnras/stt428},
  \href {http://adsabs.harvard.edu/abs/2013MNRAS.432..176P} {432, 176}

\bibitem[\protect\citeauthoryear{{Pakmor}, {Bauer}  \& {Springel}}{{Pakmor}
  et~al.}{2011}]{2011MNRAS.418.1392P}
{Pakmor} R.,  {Bauer} A.,   {Springel} V.,  2011, \mn@doi [\mnras]
  {10.1111/j.1365-2966.2011.19591.x}, \href
  {http://adsabs.harvard.edu/abs/2011MNRAS.418.1392P} {418, 1392}

\bibitem[\protect\citeauthoryear{{Pakmor}, {Springel}, {Bauer}, {Mocz},
  {Munoz}, {Ohlmann}, {Schaal}  \& {Zhu}}{{Pakmor}
  et~al.}{2016}]{2016MNRAS.455.1134P}
{Pakmor} R.,  {Springel} V.,  {Bauer} A.,  {Mocz} P.,  {Munoz} D.~J.,
  {Ohlmann} S.~T.,  {Schaal} K.,   {Zhu} C.,  2016, \mn@doi [\mnras]
  {10.1093/mnras/stv2380}, \href
  {http://adsabs.harvard.edu/abs/2016MNRAS.455.1134P} {455, 1134}

\bibitem[\protect\citeauthoryear{{Park}, {Caprioli}  \& {Spitkovsky}}{{Park}
  et~al.}{2015}]{2015PhRvL.114h5003P}
{Park} J.,  {Caprioli} D.,   {Spitkovsky} A.,  2015, \mn@doi [Physical Review
  Letters] {10.1103/PhysRevLett.114.085003}, \href
  {http://adsabs.harvard.edu/abs/2015PhRvL.114h5003P} {114, 085003}

\bibitem[\protect\citeauthoryear{{Pfrommer}, {Pakmor}, {Schaal}, {Simpson}  \&
  {Springel}}{{Pfrommer} et~al.}{2017a}]{2017MNRAS.465.4500P}
{Pfrommer} C.,  {Pakmor} R.,  {Schaal} K.,  {Simpson} C.~M.,   {Springel} V.,
  2017a, \mn@doi [\mnras] {10.1093/mnras/stw2941}, \href
  {http://adsabs.harvard.edu/abs/2017MNRAS.465.4500P} {465, 4500}

\bibitem[\protect\citeauthoryear{{Pfrommer}, {Pakmor}, {Simpson}  \&
  {Springel}}{{Pfrommer} et~al.}{2017b}]{2017ApJ...847L..13P}
{Pfrommer} C.,  {Pakmor} R.,  {Simpson} C.~M.,   {Springel} V.,  2017b, \mn@doi
  [\apjl] {10.3847/2041-8213/aa8bb1}, \href
  {http://adsabs.harvard.edu/abs/2017ApJ...847L..13P} {847, L13}

\bibitem[\protect\citeauthoryear{{Powell}, {Roe}, {Linde}, {Gombosi}  \& {De
  Zeeuw}}{{Powell} et~al.}{1999}]{1999JCoPh.154..284P}
{Powell} K.~G.,  {Roe} P.~L.,  {Linde} T.~J.,  {Gombosi} T.~I.,   {De Zeeuw}
  D.~L.,  1999, \mn@doi [J. Comput. Phys.] {10.1006/jcph.1999.6299}, \href
  {http://adsabs.harvard.edu/abs/1999JCoPh.154..284P} {154, 284}

\bibitem[\protect\citeauthoryear{{Reynolds}}{{Reynolds}}{2008}]{2008ARA&A..46...89R}
{Reynolds} S.~P.,  2008, \mn@doi [\araa]
  {10.1146/annurev.astro.46.060407.145237}, \href
  {http://adsabs.harvard.edu/abs/2008ARA%26A..46...89R} {46, 89}

\bibitem[\protect\citeauthoryear{{Sarkar}, {Nath}  \& {Sharma}}{{Sarkar}
  et~al.}{2015}]{2015MNRAS.453.3827S}
{Sarkar} K.~C.,  {Nath} B.~B.,   {Sharma} P.,  2015, \mn@doi [\mnras]
  {10.1093/mnras/stv1806}, \href
  {http://adsabs.harvard.edu/abs/2015MNRAS.453.3827S} {453, 3827}

\bibitem[\protect\citeauthoryear{{Schaal} \& {Springel}}{{Schaal} \&
  {Springel}}{2015}]{2015MNRAS.446.3992S}
{Schaal} K.,  {Springel} V.,  2015, \mn@doi [\mnras] {10.1093/mnras/stu2386},
  \href {http://adsabs.harvard.edu/abs/2015MNRAS.446.3992S} {446, 3992}

\bibitem[\protect\citeauthoryear{{Sedov}}{{Sedov}}{1959}]{1959sdmm.book.....S}
{Sedov} L.~I.,  1959, {Similarity and Dimensional Methods in Mechanics}

\bibitem[\protect\citeauthoryear{{Springel}}{{Springel}}{2010}]{2010MNRAS.401..791S}
{Springel} V.,  2010, \mn@doi [\mnras] {10.1111/j.1365-2966.2009.15715.x},
  \href {http://adsabs.harvard.edu/abs/2010MNRAS.401..791S} {401, 791}

\bibitem[\protect\citeauthoryear{{Taylor}}{{Taylor}}{1950}]{1950RSPSA.201..159T}
{Taylor} G.,  1950, \mn@doi [Proceedings of the Royal Society of London Series
  A] {10.1098/rspa.1950.0049}, \href
  {http://adsabs.harvard.edu/abs/1950RSPSA.201..159T} {201, 159}

\bibitem[\protect\citeauthoryear{{Yang}, {de O{\~n}a Wilhelmi}  \&
  {Aharonian}}{{Yang} et~al.}{2017}]{2017arXiv171002803Y}
{Yang} R.-z.,  {de O{\~n}a Wilhelmi} E.,   {Aharonian} F.,  2017, preprint,
  \href {http://adsabs.harvard.edu/abs/2017arXiv171002803Y} {} (\mn@eprint
  {arXiv} {1710.02803})

\bibitem[\protect\citeauthoryear{{van Marle}, {Meliani}  \& {Marcowith}}{{van
  Marle} et~al.}{2015}]{2015A&A...584A..49V}
{van Marle} A.~J.,  {Meliani} Z.,   {Marcowith} A.,  2015, \mn@doi [\aap]
  {10.1051/0004-6361/201425230}, \href
  {http://adsabs.harvard.edu/abs/2015A%26A...584A..49V} {584, A49}

\makeatother
\end{thebibliography}

\newpage

\appendix

\section{Convergence tests}
\label{sec:convergence}

\begin{figure}
\centering
\includegraphics[scale=0.5]{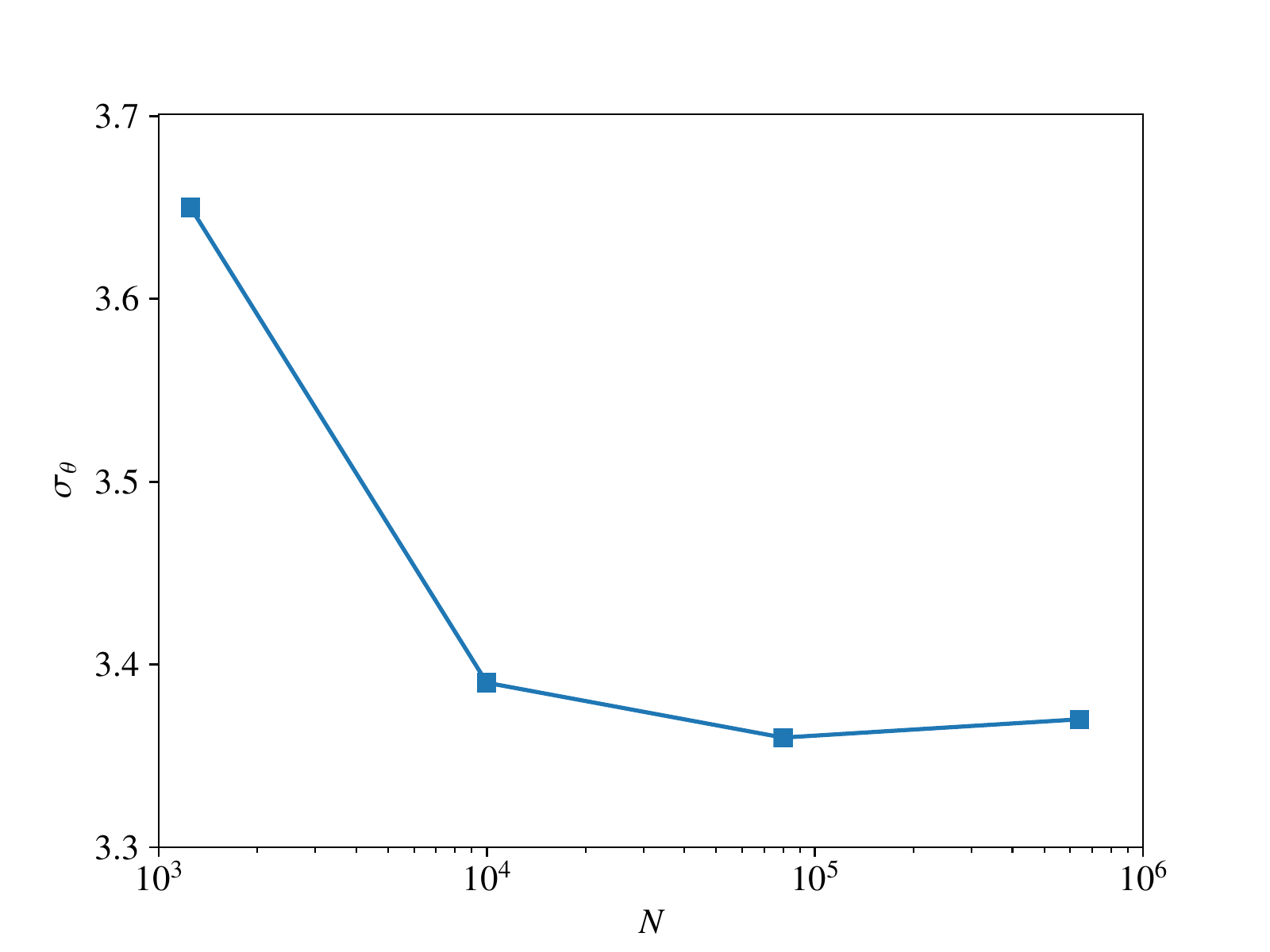}
\caption{Convergence test of the shock-tube simulation for four different
  resolutions (1200, $10^4$, $8\times 10^4$ and $64 \times 10^4$ cells) in a $10 \times 1 \times 1$
  simulation box for a magnetic obliquity of $45^\circ$. The plot shows the
  standard deviation $\sigma_{\theta}$ of the magnetic obliquity vs. the number of
  cells $N$ used in the initial setup.}
\label{figC1:convergence_shocktube}
\end{figure}

\begin{figure*}
\centering
\includegraphics[scale=1]{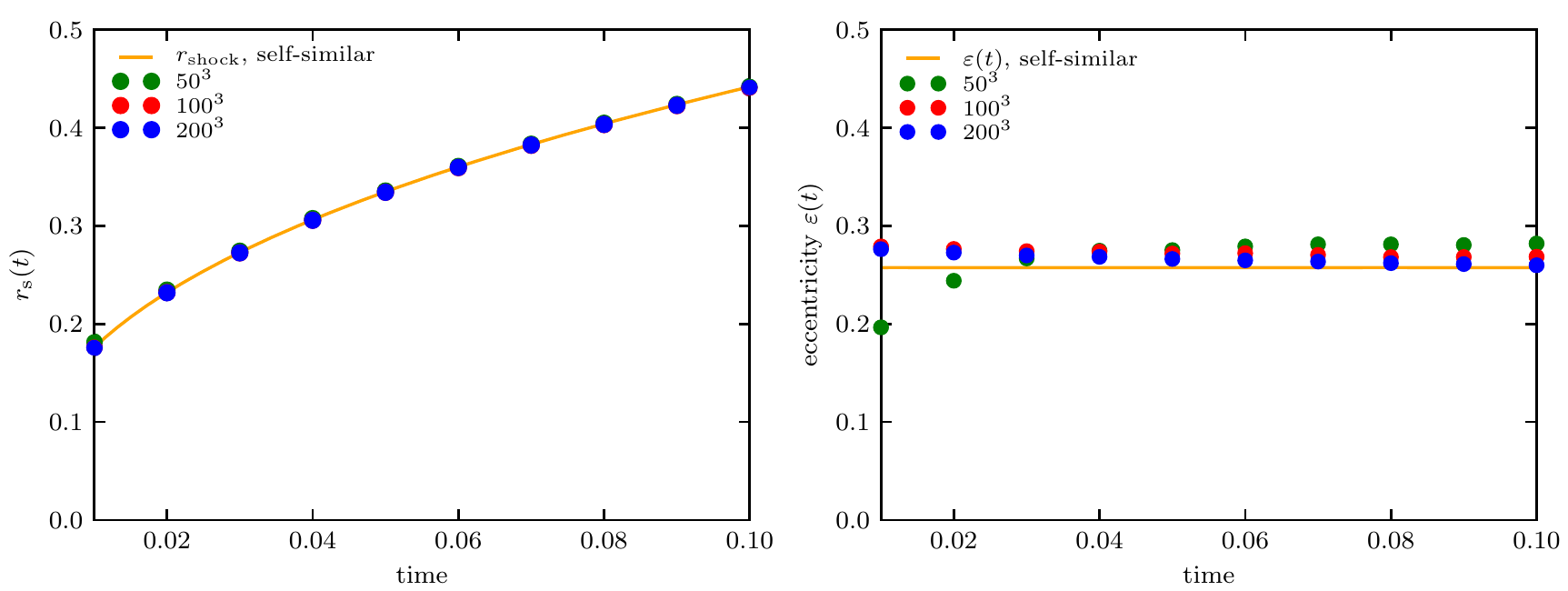}
\caption{Convergence test for the Sedov-Taylor blast wave in the case of a homogeneous
  magnetic field and obliquity dependent CR acceleration with three different
  grid resolutions: $50^3, 100^3$ and $200^3$. The left panel shows the time
  evolution of the average shock radius while the right panel shows the time
  evolution of the eccentricity of the oblate explosion.  Except for early times
  ($t<0.02$) the radius already converges for a simulation with $50^3$ cells. In
  contrast, the eccentricity converges only at a resolution of $200^3$ grid
  cells.}
\label{figC2:convergence_sedov}
\end{figure*}

Here, we perform numerical convergence tests of our shock-tube and the
Sedov-Taylor setups with a homogeneous magnetic field (see
Sections~\ref{sec:method} and \ref{sec:sedov}, respectively).  First, we asses
the convergence of the accuracy with which we recover the magnetic obliquity in
our simulations. We use several simulation outputs to measure the obliquity
distribution, which follows a Gaussian, independent of resolution.
Figure~\ref{figC1:convergence_shocktube} shows the Gaussian standard deviation
of the obliquity as a function of grid resolution, featuring 1200, $10^4$, $8
\times 10^4$ and $64 \times 10^4$ cells in our elongated shock tube setup
($10\times1\times1$).  We notice that the standard deviation $\sigma_\theta$
decreases from $1200$ to $10^4$ cells and levels off for better resolved
simulations, indicating convergence for measuring the magnetic obliquity for at
least $10^4$ cells or equivalently $10^3$ cells per individual three-dimensional
unit.  To assess the numerical convergence of our ellipsoidal Sedov-Taylor
problems with obliquity dependent CR acceleration, we perform simulations with
$50^3, 100^3 $ and $200^3$ grid cells. The results are reported in
Fig.~\ref{figC2:convergence_sedov}.  The time evolution of the average shock
radius (shown in the left panel) already converges for a $50^3$ simulation
except for the first two points. We derive the radius of our self-similar
solution with equation~(\ref{eq:alpha_final}) using an average efficiency value
taken from the $200^3$ simulation. Contrarily, the time evolution of the
eccentricity of the oblate explosion (shown in the right panel) converges much
slower and converges on our theoretical eccentricity at a resolution of $200^3$
cells. The self-similar solution of the eccentricity is constructed by fitting a
Sedov-Taylor solution of the shock evolution to the data of the $200^3$
simulation in the parallel and perpendicular regions (as defined in
Sect.~\ref{sec:homogeneous}).

\section{Details of the Sedov-Taylor solution}
\label{sec:ST_analytics}

The Sedov-Taylor similarity solution makes two fundamental assumptions: (i) it
assumes that the explosion was sufficiently long ago so that the initial
conditions do not impact the solution and (ii) that the explosion expands into a
medium of negligible pressure (or temperature). For these assumptions, the
solution describes a strong spherical shock wave whose position only depends on
the injected energy and the density of the ambient medium
\citep{1959sdmm.book.....S,1950RSPSA.201..159T}. These assumptions still hold
when including a magnetic field that is flux-frozen into the gas. We follow the
derivation by \citet{1966hydr.book.....L} and only state the starting point and
relevant definitions that are necessary to understand our final novel analytical
expression of the self-similar parameter $\alpha$ in equation~(\ref{eq:ST}).

The velocity of the shock wave relative to the background gas at rest is given
by (equation~\ref{eq:ST})
\begin{equation}
  u_1 = \dfrac{\de r_{\rmn{s}} (t)}{\de t} = \dfrac{2r_{\rmn{s}} (t)}{5t} =
  \dfrac{2}{5} \left(\dfrac{E_1}{\alpha \rho_1 t^3}\right)^{1/5}.
\end{equation}
Using the Rankine-Hugoniot expressions in the limit of strong shocks, the gas
pressure $P_2$, mass density $\rho_2$ and velocity $v_2 = u_1 - u_2$ in the
post-shock rest frame can be expressed in terms of the shock velocity $u_1$:
\begin{eqnarray}
\label{eq:v2}
v_2 &=& \dfrac{2 u_1}{\gamma+1} ,\\
\label{eq:rho2}
\rho_2 &=& \dfrac{\gamma+1}{\gamma-1} \rho_1,\\
\label{eq:P2}
P_2 &=& \dfrac{2 \rho_1 u_1^2}{\gamma+1}.
\end{eqnarray}
To determine the gas flow in the region behind the shock, we introduce dimensionless
variables $V,G,Z$ for the gas velocity $v$, density $\rho$ and the squared sound
velocity $c^2$, respectively:
\begin{eqnarray}
\label{eq:vV}
v &=& \dfrac{2r}{5t}\,V  ,\\
\label{eq:rG}
  \rho &=& \rho_1 G  ,\\
\label{eq:PZ}
c^2 &=& \dfrac{4r^2}{25 t^2}\,Z .
\end{eqnarray}
These parameters are functions of the dimensionless variable
\begin{eqnarray}
\xi = \dfrac{r}{r_{\rmn{s}}(t)} = r \left(\dfrac{\alpha \rho_1}{E_1 t^2} \right)^{1/5} .
\end{eqnarray}

Using these dimensionless quantities the conservation of energy can be expressed
in terms of $Z$ and as an implicit function of $\xi$ through $V(\xi)$
\citep{1966hydr.book.....L}:
\begin{equation}
Z = \dfrac{\gamma(\gamma-1)(1-V)V^2}{2(\gamma V -1)} .
\end{equation}
Following \citet{1966hydr.book.....L}, we arrive at the following set of
equations:
\begin{equation}
\begin{aligned}
  \xi^5 = \left[ \dfrac{1}{2}(\gamma+1)V \right]^{-2}
  &\left\lbrace \dfrac{\gamma+1}{7-\gamma}[5-(3\gamma-1)V] \right\rbrace^{\nu_1}  \\
  & \times \left[ \dfrac{\gamma+1}{\gamma-1} (\gamma V -1) \right]^{\nu_2},
 \end{aligned}
\end{equation} 
\begin{equation}
\begin{aligned}
  G = \dfrac{\gamma+1}{\gamma-1} \left[ \dfrac{\gamma+1}{\gamma-1} (\gamma V -1) \right]^{\nu_3}
  & \left\lbrace \dfrac{\gamma+1}{7-\gamma}[5-(3\gamma-1)V] \right\rbrace^{\nu_4} \\
  & \times \left[ \dfrac{\gamma+1}{\gamma-1} (1- V) \right]^{\nu_5}
\end{aligned}
\end{equation}
with
\begin{eqnarray}
  \nu_1 &=& -\dfrac{13 \gamma^2 - 7\gamma+12}{(3\gamma-1)(2\gamma+1)},\\
  \nu_2 &=& \dfrac{5(\gamma-1)}{2\gamma+1},  \\
  \nu_3 &=& \dfrac{3}{2\gamma+1},\\
  \nu_4 &=& -\dfrac{\nu_1}{2-\gamma},\\
  \nu_5 &=& -\dfrac{2}{2-\gamma}.
\end{eqnarray}

\begin{figure}
\centering
\includegraphics[scale=0.55]{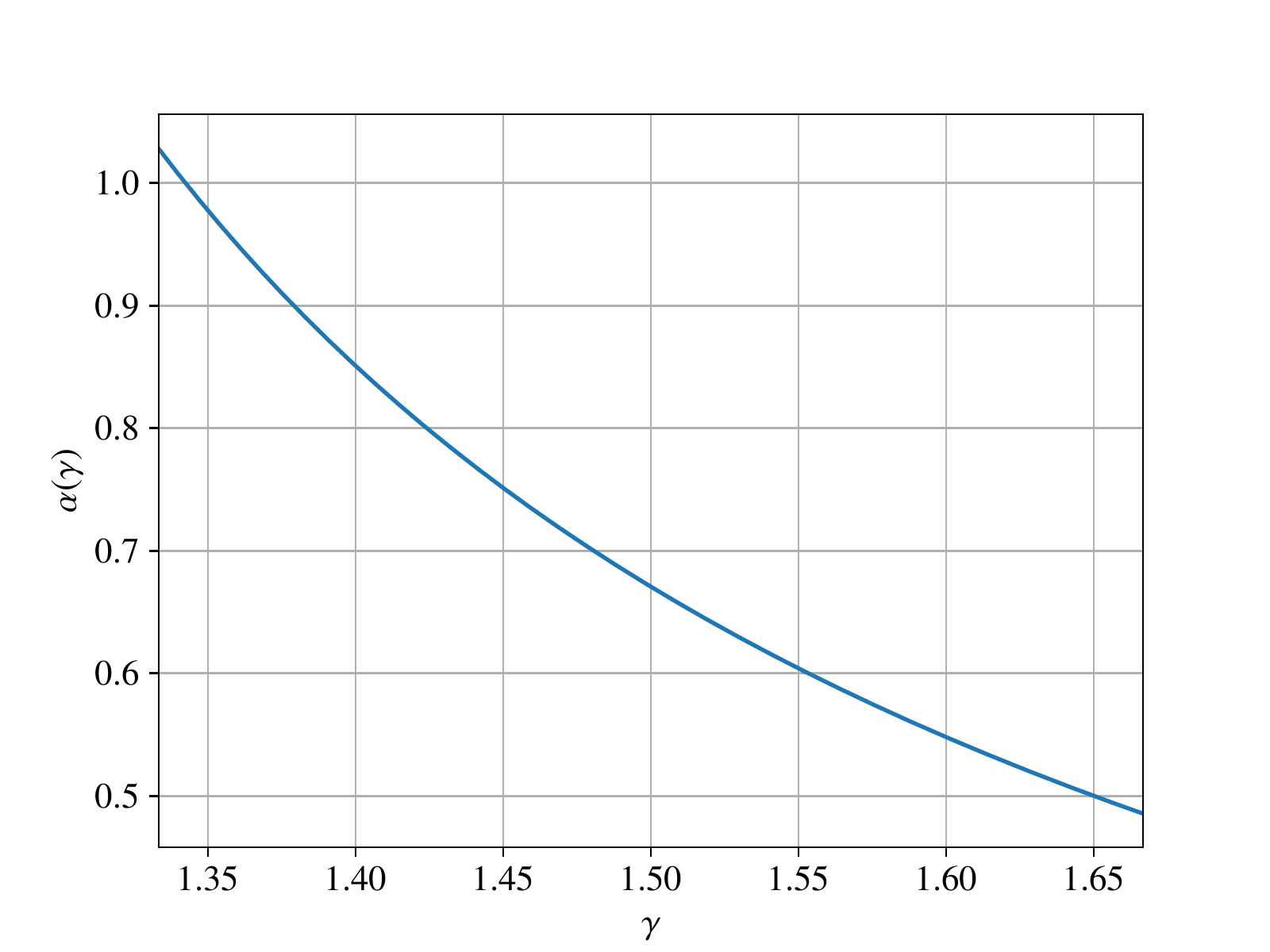}
\caption{We show the self-similarity factor $\alpha$ of the Sedov-Taylor
  solution as a function of the ratio of specific heats $\gamma$.}
\label{fig:alpha}
\end{figure}

The variable $\alpha$ as a function of the independent variable $\xi$ is
determined by the condition
\begin{equation}
  E_1 = \int_0^{r_{\rmn{s}}} \rho
  \left( \dfrac{1}{2} v^2 + \dfrac{1}{\gamma -1}\dfrac{P}{\rho}\right) 4 \pi r^2 \de r  ,
\end{equation}
which states that the total energy of the gas is equal to the released energy of
the original explosion.  In terms of dimensionless quantities, this
equation reads
\begin{equation}
\alpha = \dfrac{16}{25}\pi \int_0^1 G(\xi) \left[ \dfrac{1}{2} V^2(\xi) + \dfrac{Z(\xi)}{\gamma(\gamma-1)} \right] \xi^4 \de \xi = \alpha(\gamma)   .
\end{equation}

As $\gamma = c_P/c_V$ (where $c_P$ and $c_V$ are the specific heats at
constant volume and pressure, respectively) we have $1 < \gamma < 2$.  In our
simulations we adopt values of $\gamma$ in the range $[4/3, 5/3]$, such that
$\alpha(\gamma)$ can be approximated with high precision, slightly modifying the
formula used by \citet{1984oup..book.....M}:
\begin{equation}
  \alpha(\gamma) \approx \dfrac{16}{75}
  \left[ \dfrac{\pi (3\gamma-1)}{(\gamma-1)(\gamma+1)^2} -\dfrac{3}{8}\right]  ,
\end{equation} 
which is accurate to within 0.8\% and is shown in Fig.~\ref{fig:alpha}.

The dimensionless quantities defined via equations~\eqref{eq:vV}, \eqref{eq:rG}
and \eqref{eq:PZ} yield the implicit expressions for $v(r), \rho(r)$ and $P(r)$:
\begin{eqnarray}
v(r) &=& \dfrac{1}{2} (\gamma +1 ) r V(r) ,\\
\rho(r) &=& G(r) \rho_1 ,\\
P(r) &= & \dfrac{2\rho_1 u^2_1 }{\gamma+1} \left[ \dfrac{1}{2} (\gamma + 1) V\right]^{-6/5} \left[ \dfrac{\gamma + 1}{\gamma - 1} \left( 1 -  V \right) \right]^{-\nu_5 + 1} \nonumber\\
&& \times \left\lbrace \dfrac{\gamma+1}{7-\gamma}[5-(3\gamma-1)V] \right\rbrace^{-\frac{\nu_4+2\nu_1}{5}}.
\end{eqnarray}

\section{Ellipsoidal reference frame} 
\label{sec:ellipsoid}

\begin{figure}
\centering
\includegraphics[scale=0.5]{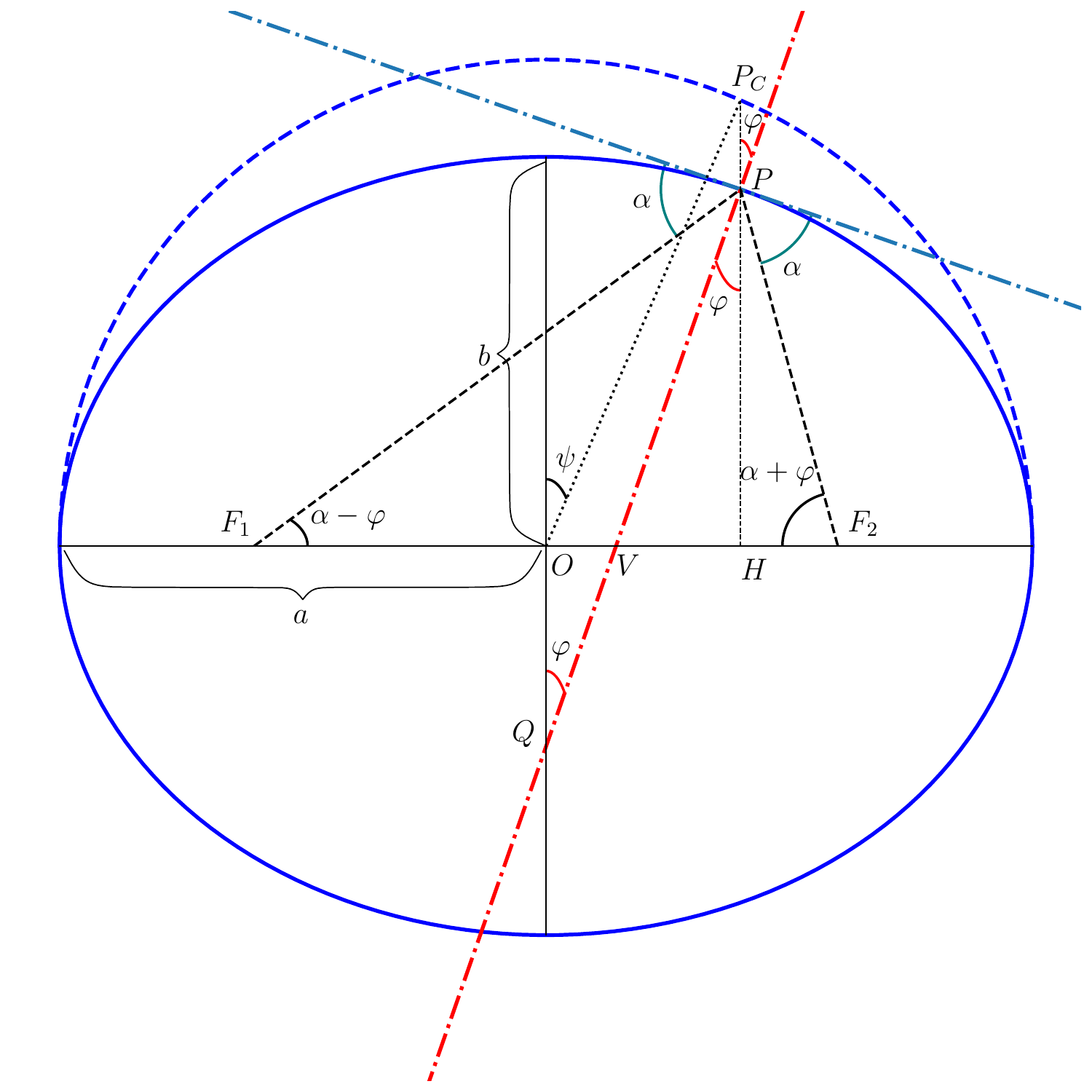}
\caption{Elliptical section of an oblate that defines the ellipsoidal reference
  frame with the pseudo-azimuthal angle $\varphi$. The two focal points $F_1$
  and $F_2$ determine a point $P$ on the surface of the ellipse. The tangent to
  that point (blue dash-dotted line) forms two equal angles $\alpha$ with the
  lines to the focii $\overline{P F_1}$ and $\overline{P F_2}$. Thus, the
  perpendicular to the tangent in $P$ (red dash-dotted line) intersects the $z$
  axis in point $Q$, which varies with the position of $P$ and forms the desired
  angle $\varphi$. }
\label{figD1:ellipse}
\end{figure}

The radial unit vector of a spherical coordinate system is not perpendicular to
an oblate surface except for the poles at $z =\pm b$, which would complicate the
relation to the magnetic obliquity for a homogeneous magnetic field aligned with
the $z$ axis. To simplify our computation of the magnetic obliquity on an oblate
surface, we adopt the ellipsoid coordinate system. Here, the bisector of a
tangent to point $P$ intersects the $z$ axis in $Q$, which varies according to
the position of the point $P$ on the oblate surface (see
Fig.~\ref{figD1:ellipse}). It assumes values from $0$ (for a point on the
semi-major axis) to $-\infty$ (for a point on the semi-minor axis).

A point $P$ on the ellipse has the property that the sum of distances from the
two focal points $F_1$ and $F_2$ to $P$ is constant. A tangent to the ellipse in
that point forms equal angles $\alpha$ with the two focal segments
$\overline{F_1 P}$ and $\overline{F_2 P}$. Dropping the perpendicular to the
tangent in $P$, by construction bisects the angle and intersects the $z$ axis in
$Q$. The angle between this bisector and the $z$ axis defines the angle
$\varphi$. Assuming a homogeneous magnetic field that is aligned with the $z$
axis, implies that the pseudo-azimuthal angle coincides with the magnetic
obliquity, $\varphi=\theta$.

The point $P$ can be vertically projected onto a circumference in the point
$P_C$, which forms an azimuthal angle $\psi$ with respect to the semi-minor axis
$b$. The angle $\varphi$ is related to this angle via the following formula:
\begin{equation}
\tan \varphi = \dfrac{b}{a} \tan \psi   .
\end{equation}

\section{Sedov-Taylor solution of a dipole field}
\label{sec:dipole}

\begin{figure*}
\centering
\includegraphics[scale=1.0]{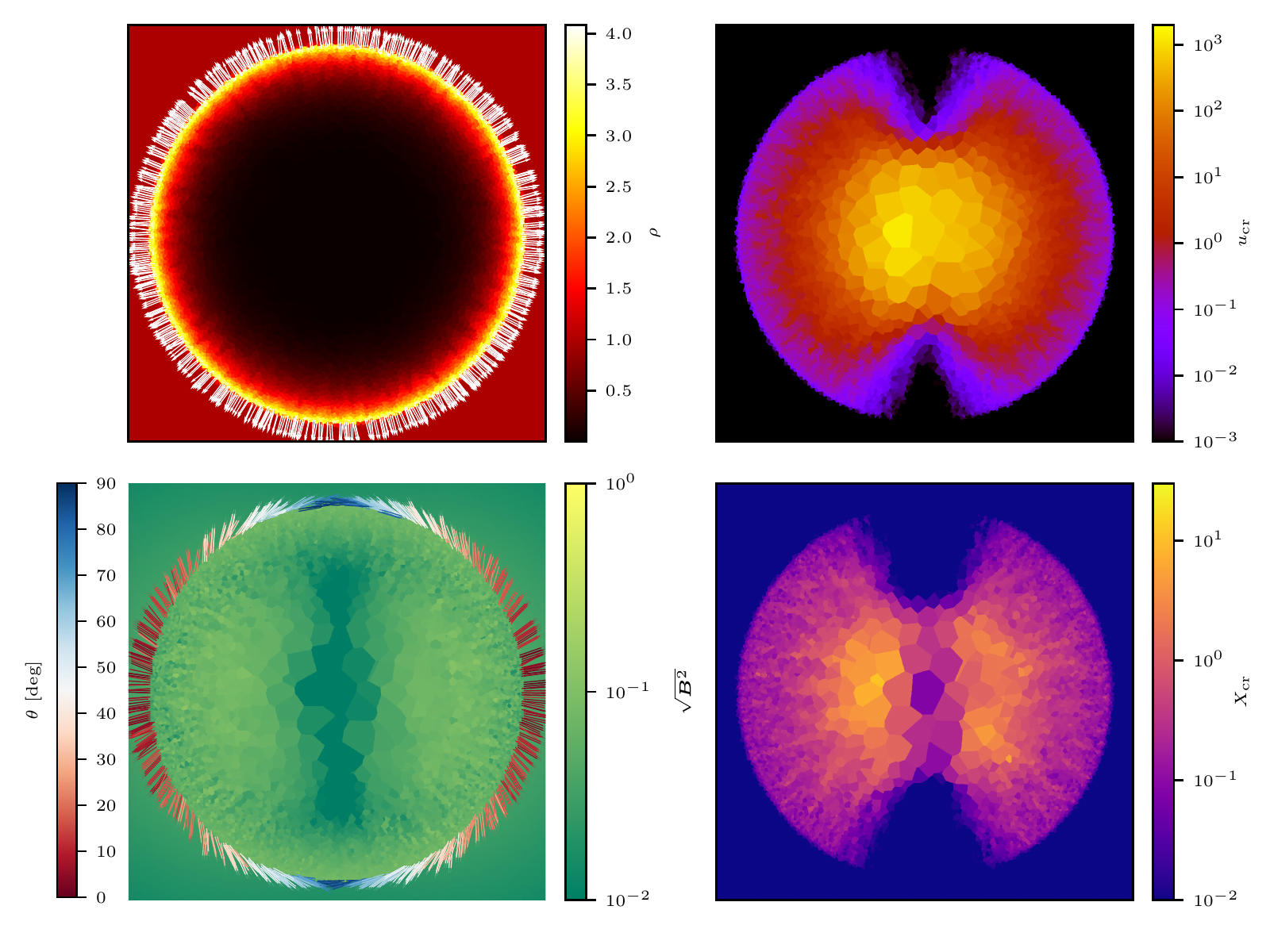}
\caption{Sedov-Taylor blast wave with obliquity dependent CR acceleration
  expanding into a {\it dipole magnetic field} that is centered at point of the
  explosion and initially oriented horizontally. Quantities are shown at $t=0.1$
  and are the same as in Fig.~\ref{fig5:sedov_maps}.  Despite the different
  magnetic field morphology in comparison to the homogeneous case, the specific
  CR energy still exhibits a quadrupolar anisotropy, but with a broader region
  of CR production.}
\label{figB1:dipole}
\end{figure*}

\begin{figure*}
\centering
\includegraphics[scale=0.7]{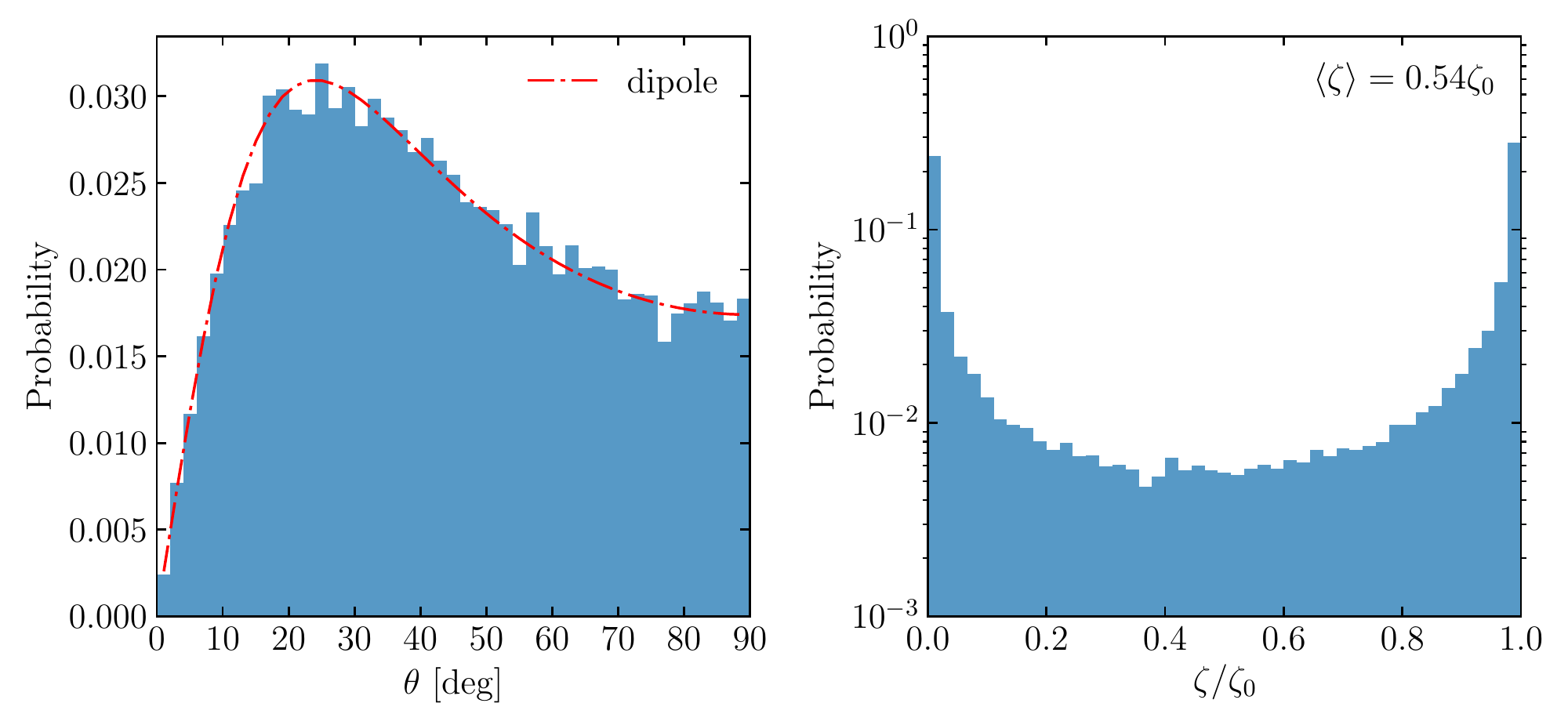}
\caption{Probability distribution functions of the magnetic obliquity for the
  magnetic dipole case (left) and the resulting bimodal distribution for the
  acceleration efficiency (right). The theoretical distribution $f(\theta)$ of
  equation~\eqref{eq. B7:Dtheta} (red dot-dashed line) compares nicely to our
  simulations.}
\label{figB2:histogram_dipole}
\end{figure*}

As a last application, we study CR shock acceleration in the case of a magnetic
dipole field with the dipole moment pointing in the positive $x$ direction. We
chose the dipole configuration because it is expected to be the dominant
magnetic configuration emerging from a non-rotating star at large distances and
because of its self-similarity.  While the magnitude of the magnetic field
strength decreases as $r^{-3}$, where $r$ is distance from the source, the
dipole field shows a constant magnetic obliquity at fixed latitude.  This
implies that the explosion encounters exactly the same replica of the magnetic
field at different radii. We perform a $100^3$-cell simulation with an extremely
low $\zeta_0 = 0.02$ to reduce the effect of CR pressure on the explosion shape
and apply a Plummer-type softening length for $r \rightarrow 0$ to avoid
magnetic divergence at the origin.  The magnetic field in the polar regions is
oriented mostly parallel to the shock normal, which results in efficient CR
acceleration as the blast wave sweeps across it. The resulting quadrupolar CR
morphology is shown in Fig.~\ref{figB1:dipole}, resembling qualitative
similarities to the homogeneous field case.

To analytically calculate the average efficiency we proceeded as follows. The
normalised radial component of the magnetic dipole field is given by
\begin{equation}
\bm{\hat{b}} \bcdot \bm{\hat{r}} = \dfrac{2 \cos \vartheta}{\sqrt{1 + 3\cos^2\vartheta}} ,
\end{equation}
where $\bm{\hat{r}}$ denotes a radial unit vector and $\vartheta$ is the
azimuthal angle. Thus, the magnetic obliquity reads as
\begin{equation}
\label{eq. B3: thetaphi}
\theta(\vartheta) = \arccos(\bm{\hat{b}} \bcdot \bm{\hat{r}})
=\arccos\left( \dfrac{2 \cos\vartheta}{\sqrt{1 + 3 \cos^2\vartheta}} \right) .
\end{equation}
Capitalizing on the symmetry of both hemispheres, we calculate the average
efficiency, 
\begin{equation}
\langle \zeta \rangle = \int_0^{\pi/2} \zeta[\theta(\vartheta)] \sin \vartheta \ \de \vartheta = 0.55 \zeta_0 .
\end{equation}
Note that the result is considerably larger in comparison to our previously discussed 
case of CR acceleration in a turbulent field.

In Fig.~\ref{figB2:histogram_dipole} we show the obliquity distribution for a
dipole field. To analytically predict this distribution, we invert
equation~\eqref{eq. B3: thetaphi} and obtain
\begin{equation}
  \vartheta (\theta) = \arccos\left( \dfrac{\cos\theta}{\sqrt{4 - 3 \cos^2\theta}} \right),
  \mbox{ with } \theta\in[0,\pi/2].
\end{equation}
The distribution of magnetic obliquity in the case of a dipole field is obtained
through a change of variables:
\begin{equation}
  \sin \vartheta  \ \de \vartheta
  = \sin[\vartheta(\theta)] \left(\dfrac{\de \vartheta}{\de \theta}\right) \de \theta
  = f(\theta) \de  \theta ,
\end{equation}
where 
\begin{equation}
\label{eq. B7:Dtheta}
f(\theta)  = \dfrac{4 \sin\theta}{(4-3\cos\theta)^{3/2}} .
\end{equation}
This analytical result compares favorably to the simulations (see left-hand
panel of Fig.~\ref{figB2:histogram_dipole}). The simulated and theoretically
expected average acceleration efficiencies agree within $2\%$ (for our $100^3$
cell simulation), demonstrating the accuracy of our numerical algorithms.

\end{document}